\RequirePackage{amsmath}
\documentclass[a4paper,fleqn,usenatbib]{mnras}

\usepackage{mathptmx}
\usepackage{txfonts}

\usepackage[T1]{fontenc}
\usepackage{ae,aecompl}

\usepackage{graphicx}	
\usepackage{amsmath}	
\usepackage{amssymb}	
\usepackage{subfig}
\usepackage{bm}
\usepackage{rotate}
\usepackage{rotating}
\usepackage{verbatim}
\usepackage{hyperref}
\newif\ifAMStwofonts
\AMStwofontstrue

\def\aj{\rmfamily{AJ~}}           
\def\apj{\rmfamily{ApJ~}}         
\def\apjl{\rmfamily{ApJ~}}        
\def\apjs{\rmfamily{ApJS~}}       

\def\aap{\rmfamily{A\&A~}}        

\def\mnras{\rmfamily{MNRAS~}}     
\def\nat{\rmfamily{Nature~}}      
\def\ao{\rmfamily{Appl.~Opt.~}}   

\def\gs{\mathrel{\hbox{\rlap{\hbox{\lower4pt\hbox{$\sim$}}}\hbox{$>$}}}}
\def\ls{\mathrel{\hbox{\rlap{\hbox{\lower4pt\hbox{$\sim$}}}\hbox{$<$}}}}

\def\asca{{\it ASCA~}}

\def\xmm{{\it XMM-Newton~}}

\def\mrk335{{Mrk~335}}

\def\ga{{\rm\thinspace gauss}}

\def\src{{IRAS 17020+4544~}}

\title[ Continuum variability in \src ]{ Characterising continuum variability in the radio-loud narrow-line Seyfert 1 galaxy \src }

\author[A. G. Gonzalez et al.]{A.~G. Gonzalez$^{1}$\thanks{E-mail: agonzalez@ap.smu.ca},
			L.~C. Gallo$^{1}$,
			P. Kosec$^{2}$,
			A.~C. Fabian$^{2}$,
			W.~N. Alston$^{2}$,
\newauthor
			M. Berton$^{3,4}$,
			and D.~R. Wilkins$^{5}$\thanks{Einstein Fellow}
\\
			$^{1}$Department of Astronomy and Physics, Saint Mary's University, 923 Robie Street, Halifax, NS, B3H 3C3, Canada\\
			$^{2}$Institute of Astronomy, Madingley Road, Cambridge, CB3 OHA, UK\\
			$^{3}$Finnish Centre for Astronomy with ESO (FINCA), University of Turku, Quantum, Vesilinnantie 5, FI-20014 University of Turku, \\Finland\\
			$^{4}$Aalto University Mets{\"a}hovi Radio Observatory, Mets{\"a}hovintie 114, FI-02540 Kylm{\"a}l{\"a} Finland\\
			$^{5}$Kavli Institute for Particle Astrophysics and Cosmology, Stanford University, 452 Lomita Mall, Stanford, CA 94305, USA
}

\date{Accepted 2020 June 12. Received 2020 May 9; in original form 2020 February 19}

\pubyear{2020}

\begin{document}
\label{firstpage}
\pagerange{\pageref{firstpage}--\pageref{lastpage}}
\maketitle

\begin{abstract}
We present results of temporal and spectral analyses on four \xmm EPIC pn observations of IRAS 17020+4544, a narrow-line Seyfert 1 galaxy with evidence of a radio jet. Analysis of the light curves reveals that this radio-loud source does not behave like the bulk population of its radio-quiet counterparts. A trend of spectral hardening with increased flux is found. Variability is found to increase with energy, though it decreases as the spectrum hardens. The first $40~\mathrm{ks}$ of the most recent observation behave uniquely among the epochs, exhibiting a softer spectral state than at any other time. Possible non-stationarity at low energies is found, with no such effect present at higher energies, suggesting at least two distinct spectral components. A reverberation signature is confirmed, with the lag-frequency, lag-energy, and covariance spectra changing significantly during the soft-state epoch. The temporal analysis suggests a variable power-law in the presence of a reflection component, thus motivating such a fit for the $0.3-10~\mathrm{keV}$ EPIC pn spectra from all epochs. We find an acceptable spectral fit using the timing-motivated parameters and report the detection of a broad Fe K emission line, requiring an additional model component beyond the reflection spectrum. We discuss links between this source and other narrow-line Seyfert 1 sources that show evidence of jet activity, finding similarities among this currently very limited sample of interesting objects.
\end{abstract}




\section{Introduction}
\label{sect:introduction}
Evidence suggests that every massive galaxy likely harbours a super-massive black hole (SMBH) at its centre. Active galactic nuclei (AGN) are those galaxies where the SMBH is actively accreting material from the surrounding environment. This accretion process may result in the formation of an accretion disc \citep{ShakuraSunyaev1973}, enabling maximally efficient conversion of matter into radiation which thereby produces the most luminous continuous sources of emission in the Universe that in many cases outshines the entire stellar component of the AGN host galaxy. 

AGN are luminous emitters across the entire electromagnetic spectrum, exhibiting significant variability in flux and spectral state over time. In the X-ray band AGN can be broadly described as having a two-component spectral model, dominated by a blackbody-like component at lower energies (i.e. $\lesssim1~\mathrm{keV}$) and a power-law component over a broader high-energy band (i.e. $>2~\mathrm{keV}$). This simple model is often modified significantly, such as by absorption in the host galaxy and/or contributions from X-ray reflection off of the accretion disc, greatly complicating the inferences that can be made about the physical environment surrounding the SMBH when modelling the X-ray spectrum. In order to alleviate some of these modelling complications various model-independent techniques have been brought to the fore-front of AGN research, with notable success. An important caveat of such techniques, however, is the need for large quantities of high quality data, sometimes limiting their use to well-sampled, bright sources.

Narrow-line Seyfert 1 galaxies (NLS1s) are a type of AGN that frequently exhibit extreme, complex behaviour for a number of different properties. These AGN were originally classified as those displaying H$\beta$ emission lines with full-width half-maximum values (FWHM) of $<2000~\mathrm{km~s^{-1}}$ \citep{Goodrich1989}, as well as strong Fe \textsc{ii} and weak [O \textsc{iii}] emission \citep{OsterbrockPogge1985}. The FWHM of these H$\beta$ features, originating from the broad-line region (BLR) which is dominated by orbital motion governed by the gravitational potential of the central SMBH, suggest smaller black hole masses ($M_{\mathrm{BH}}<10^8~M_{\odot}$) for the NLS1 population (e.g. \citealt{Peterson+2000}). Other optical-regime works have revealed accretion rates in these AGN to be between $L/L_{\mathrm{Edd}}\approx0.1-1$ \citep{BorosonGreen1992}, in some cases even surpassing the Eddington limit, indicating an extreme central environment. 

Radio studies of NLS1s have found that a large majority are undetected across radio wavelengths (i.e. radio-silent) or exhibit low levels of emission (i.e. radio quiet; RQ). A small fraction ($\sim7~\mathrm{per~cent}$) are defined as being radio-loud (RL) \citep{Komossa+2006} possessing powerful, relativistic jets. There is, however, some ambiguity surrounding radio-loudness as a parameter and its usefulness in characterising and / or classifying AGN as there are now examples of previously claimed radio-silent sources being detected at $37~\mathrm{GHz}$ indicating jet activity \citep{Lahteenmaki+2017,Lahteenmaki+2018}.

In the X-ray regime, NLS1s typically display steep spectra (i.e. X-ray power-law photon index $\Gamma \gtrsim 2$), the presence of a significant soft excess, and rapid, sometimes large amplitude, variability (e.g. \citealt{Boeller+1996, Brandt+1997, Komossa+2017, Gallo2018}). The steep X-ray spectrum may be indicative of a cooler or lower density primary corona than is found in other types of AGN  while the strong soft excess might arise from a large contribution of X-ray reflection off of the accretion disc. 

One of the many open questions in AGN research seeks to uncover the connection between the accretion disc and radio jet (i.e. disc-jet connection). Previous works have suggested that in states of sub-Eddington, radiatively inefficient accretion, emission is dominated by the jet and that once the source enters a higher, more efficient accretion state the disc emission begins to quench that of the jet (e.g. \citealt{Merloni+2003,Kording+2006}). Coupling the strong disc emission and evidence of jet activity in the RL-NLS1 population offers a unique test-bed in which a connection between these two components may be explored.

For instance, the structure of the X-ray power-law source (i.e. corona) in AGN has historically been represented as a compact `point-source' geometry that resides some height above the accretion disc. Recently, however, there have been results indicating signatures of coronal geometries that are low in height and significantly extended over the inner disc as well as signatures of vertically collimated coronae that are outflowing at some velocity. In Mrk 335 \citep{WilkinsGallo2015} and I Zw 1 \citep{Wilkins+2017}, two examples of RQ-NLS1s, events that can be described as aborted jet launches have been suggested, wherein an originally extended corona became vertically collimated (gaining some outflow velocity) before collapsing into a more compact geometry.

Many NLS1s also display clear signatures of emission from the inner accretion disc, such as a soft excess at low energies $\lesssim1~\mathrm{keV}$ and relativistically broadened Fe K emission in the $6-7~\mathrm{keV}$ band originating from material at the inner most edge of the disc \citep{Laor1991}. Reverberation analysis techniques have been used as a model-independent tool to reveal time-lags between the soft excess and coronal emission in numerous AGN (e.g. \citealt{Kara+2016}). In some cases a high-frequency lag is detected such that emission from the soft excess lags behind the continuum, supporting the idea of a reflection-based origin for the soft excess. Such features are also observable in the lag-energy spectra, constructed by plotting the aforementioned reverberation lags as a function of energy, where they show large time lags for the soft excess and Fe K emission band compared with the continuum, indicating a common origin (e.g. \citealt{Zoghbi+2010, Kara+2016}). In some sources, however, the soft excess has been interpreted as evidence of a secondary cooler, optically thick X-ray corona (e.g. \citealt{Done+2012}). 

RL-NLS1s therefore pose a promising population of AGN in which the disc-jet connection can be rigorously tested, owing to the recent evidence of launching X-ray coronae and strong reflected emission from the accretion disc.

\src (also known as: B3 1702+457, J170330.38+454047.1) is a nearby ($z=0.0604$)\footnote{Obtained from the NASA/IPAC Extragalactic Database (NED) at \href{https://ned.ipac.caltech.edu/}{https://ned.ipac.caltech.edu/}} RL-NLS1 that has been studied numerous times at radio wavelengths with a sparse presence in X-ray astronomy literature. Radio surveys have found the source to be classified as a compact steep-spectrum object (CSS) with a turnover frequency at $<150~\mathrm{MHz}$ \citep{Snellen+2004} exhibiting a sub-relativistic or misaligned jet on parsec scales \citep{Doi+2011,Giroletti+2017}. \cite{Berton+2018}, however, reported a flat in-band spectral index while also finding a steep broad band index, suggesting significant radio variability. Early work \citep{Leighly+1997,KomossaBade1998} discovered the presence of warm absorbers (WAs) in the X-ray spectrum, with recent studies \citep{Longinotti+2015, Sanfrutos+2018} confirming their presence alongside ultra-fast outflows of high column density, highly ionised material. The high-energy X-ray spectrum (i.e. $0.3-10~\mathrm{keV}$) has not yet been examined.

In this work we will focus on the analysis of four \xmm EPIC pn observations of \src presenting timing and spectral analyses exploring the observed variability. The paper is organised as follows. In Section \ref{sect:observations} we present details of the observations and data reduction. Section \ref{sect:timing} presents the various timing methods used to characterise the observed variability using model-independent approaches. In Section \ref{sect:spectral} we model the X-ray spectrum using spectral models motivated by the timing results. We discuss further the implications of our results in Section \ref{sect:discussion}, concluding in Section \ref{sect:conclusion}.

\section{Observations \& Data Reduction}
\label{sect:observations}

In this paper we use archival data obtained from the \xmm \citep{Jansen+2001} X-ray observatory. Details of the four observations used in this analysis are presented in Table \ref{tab:obs}. 

\begin{table}
	\begin{center}
		\caption{\xmm EPIC pn observations of \src used in this work. The columns are: (1) observation ID, (2) abbreviated name referenced throughout the manuscript, (3) start date of observation, (3) observation duration, and (4) final exposure time including background flaring corrections.}
		\begin{tabular}{ccccc}
			\hline
			(1) & (2) & (3) & (4) & (5) \\
			Obs. ID & Abbrev. & Start Date & Duration & Exp. \\
			& & [yyyy-mm-dd] & [ks] & [ks] \\
			\hline
			0206860101 & 2004.101 & 2004-08-30 & 21.8 & 13.5 \\
		    0206860201 & 2004.201 & 2004-09-05 & 21.5 & 12.9 \\
			0721220101 & 2014.101 & 2014-01-23 & 82 & 56.2 \\
			0721220301 & 2014.301 & 2014-01-25 & 84 & 56.0 \\
			\hline
			\label{tab:obs}
		\end{tabular}
	\end{center}
\end{table}

The EPIC pn \citep{Struder+2001} camera was operated in small window mode with the medium filter applied for the duration of all four observations. Throughout this work we abbreviate the observation IDs as in column (2) of Table \ref{tab:obs}. The \xmm Observation Data Files (ODFs) were processed using the \xmm Science Analysis System (SAS) version 15.0.0 creating event lists with the \textsc{epchain} task. While producing the event list products we used the conditions $0 \leq \mathrm{\textsc{pattern}} \leq 4$ and $\mathrm{\textsc{flag}} == 0$. Source photons were extracted from a 35$^{\prime\prime}$ circular region centred on the source. Background photons were extracted from a 50$^{\prime\prime}$ circular off-source region. The timing and spectral products were each extracted from the resulting event lists using the \textsc{evselect} task, finding evidence of significant high-energy ($>10~\mathrm{keV}$) background flaring in both 2004 observations and mild background activity during the 2014 observations; none of the observations displayed any evidence of pile-up according to the \textsc{epatplot} output. Corrected light curves for each observation were produced using the SAS task \textsc{epiclccorr}. Spectra for each observation were produced with response matrices generated using the SAS tasks \textsc{rmfgen} and \textsc{arfgen}. When producing the spectra we extracted good-time intervals (GTIs) using the \textsc{tabgtigen} task, excluding times where the background level at $>10~\mathrm{keV}$ exceeded $0.3~\mathrm{counts~s^{-1}}$, the `quiescent' background counts before and after the flaring. The final exposures after extracting the GTIs are listed as column (5) of Table \ref{tab:obs}.

We use a Galactic column density of $N_{\mathrm{H}} = 2.20\times10^{20}~\mathrm{cm^{-2}}$ from \cite{Willingale+2013}\footnote{Browser based tool for calculating Galactic column density available at \href{http://www.swift.ac.uk/analysis/nhtot/index.php}{http://www.swift.ac.uk/analysis/nhtot/index.php}}, which reports a combined atomic and molecular column density of hydrogen, and use \cite{Wilms+2000} abundances throughout our spectral modelling. We also assume a cosmology with $H_0 = 70~\mathrm{km~s^{-1}~Mpc^{-1}}$, $q_0 = 0$, and $\Lambda_0 = 0.73$. All reported spectral model parameter errors are at the $90~\mathrm{per~cent}$ confidence level, unless otherwise stated. All energy values are in the observed frame, unless otherwise stated.

\section{Timing Analysis}
\label{sect:timing}

\subsection{Exploring the light curves}
\begin{figure*}
	\scalebox{1.0}{\includegraphics[width=0.98\linewidth]{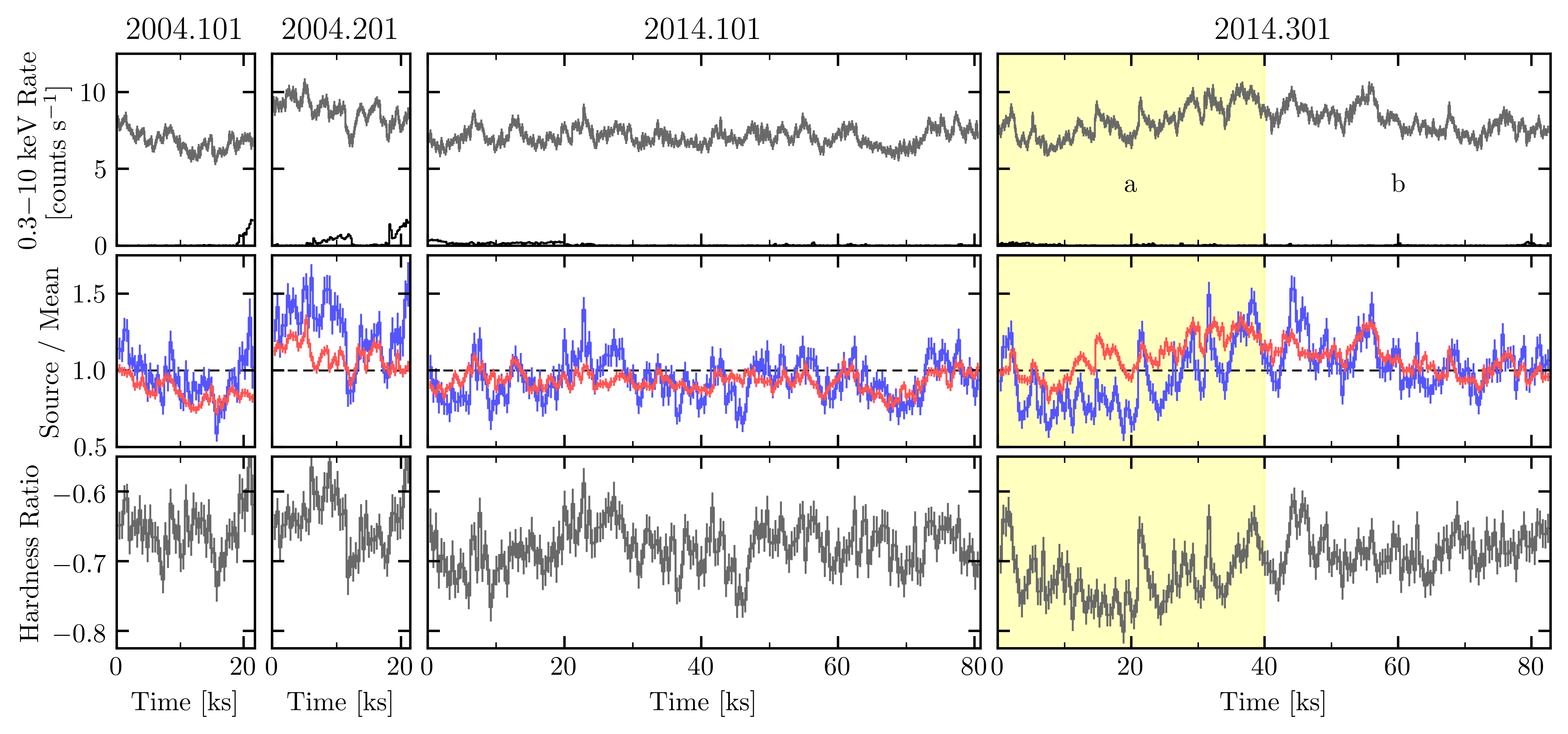}}
	\caption{\textit{Top row} $-$ The broad band $0.3 - 10~\mathrm{keV}$ EPIC pn background-subtracted light curves binned by $200~\mathrm{s}$ (grey curves) for each of the \xmm observations, with the background light curves shown as the black curves. \textit{Middle row} $-$ The soft ($S = 0.3-1~\mathrm{keV}$; red curves) and hard ($H = 2-10~\mathrm{keV}$; blue curves) EPIC pn source light curves binned by $400~\mathrm{s}$. Each light curve has been divided by the mean source count rate across all observations in the corresponding energy band allowing for direct comparisons of variability in each band. \textit{Bottom row} $-$ The hardness ratios $HR = \left(H-S\right)/\left(H+S\right)$ for each observation using the same $400~\mathrm{s}$ binned light curves as the middle row. In each of the 2014.301 panels the first $40~\mathrm{ks}$ are highlighted in yellow.}
	\label{fig:bb-lcs}
\end{figure*}

\begin{table}
	\begin{center}
		\caption{General light curve properties for each of the five epochs using time bins of $400~\mathrm{s}$. The columns are: (1) observation abbreviation, (2) the mean $0.3-10~\mathrm{keV}$ source count rate, (3) the fractional variability ($F_{\mathrm{var}}$) of the $0.3-10~\mathrm{keV}$ light curve, and (4) the mean hardness ratio ($HR = \left(H-S\right)/\left(H+S\right)$) between the $0.3-1~\mathrm{keV}$ and $2-10~\mathrm{keV}$ light curves.}
		\begin{tabular}{cccc}
			\hline
			(1) & (2) & (3) & (4) \\
			Epoch & Mean Rate & $F_{\mathrm{var}}$ & Mean $HR$ \\
			& [$\mathrm{counts~s^{-1}}$] & [$\mathrm{per~cent}$] & \\
			\hline
			2004.101 & $6.82\pm0.01$ & $8.7\pm0.3$ & $-0.655\pm0.003$ \\
			2004.201 & $8.76\pm0.02$ & $8.7\pm0.3$ & $-0.641\pm0.003$ \\
			2014.101 & $7.100\pm0.003$ & $6.9\pm0.2$ & $-0.680\pm0.001$ \\
			2014.301a & $8.11\pm0.01$ & $12.9\pm0.2$ & $-0.724\pm0.002$ \\
			2014.301b & $8.171\pm0.008$ & $9.6\pm0.2$ & $-0.685\pm0.002$ \\
			\hline
			\label{tab:lcprops}
		\end{tabular}
	\end{center}
\end{table}

For each of the four observations a broad band $0.3 - 10~\mathrm{keV}$ background-subtracted light curve was made using bins of $200~\mathrm{s}$, as shown in the top row of Figure \ref{fig:bb-lcs}. Increased levels of background activity can be seen at the end of both 2004 epochs, with 2004.201 additionally exhibiting increased background levels during the middle of the observation. The first $\sim20~\mathrm{ks}$ of 2014.101 also exhibit increased background levels. These times were filtered out using the GTI criteria described in Section \ref{sect:observations} when making the spectra for these observations. We checked and confirmed that the timing products are unaffected by these periods of increased background activity, and therefore we include them in all of the following timing analyses. The mean source count rate of each broad band light curve is shown as column (2) in Table \ref{tab:lcprops}, which globally varies within $\sim35~\mathrm{per~cent}$ of its mean of $7.7\pm1.0~\mathrm{counts~s^{-1}}$. 

The variability in the broad band light curves can be quantitatively assessed by computing the fractional variability as in \cite{Vaughan+2003}, shown as column (3) in Table \ref{tab:lcprops}. Epoch 2014.101 stands out as having the lowest fractional variability while 2014.301 displays significantly more variable behaviour than any other epoch. Both 2004 observations display a mild dimming over time while sharing similar levels of variability. These levels of variability are not extreme when compared to the large-amplitude short-term variability observed in other NLS1s.

To explore energy-dependent variability we extracted soft ($S$) and hard ($H$) band light curves using $0.3 - 1~\mathrm{keV}$ and $2 - 10~\mathrm{keV}$, respectively, and computed the hardness ratio $HR = \left(H-S\right)/\left(H+S\right)$ between them, as shown in the middle (normalised light curves) and bottom (hardness ratios) rows of Figure \ref{fig:bb-lcs}. To reduce the size of the error bars we re-binned the data into $400~\mathrm{s}$ time bins. It is clear that the hard band is significantly more variable than the soft band in all observations, having $F_{\mathrm{var},H} = 14 \pm 1$ compared to $F_{\mathrm{var},S} = 7.8 \pm 0.7$ globally. Visual inspection reveals hints of correlated variability between the soft and hard bands, to be examined further in a following section. The hardness ratios remain relatively constant throughout each observation, exhibiting variations of $\sim20~\mathrm{per~cent}$ globally, with the mean value for each epoch given as column (4) of Table \ref{tab:lcprops}. Epoch 2014.301 exhibits the most significant hardness ratio deviation of any epoch, where a dip at $\sim20~\mathrm{ks}$ is observed, after which the hardness ratio increases gradually for the remainder of the observation. 

The increased fractional variability and changing hardness ratio of 2014.301 indicate significantly different behaviour during this observation when compared to the other epochs. Since the apparent spectral state change in 2014.301 is only present during the first $40~\mathrm{ks}$ we segmented this observation into two $40~\mathrm{ks}$ long segments, hereafter referred to as 2014.301a for the first $40~\mathrm{ks}$ and 2014.301b for the remaining $40~\mathrm{ks}$. 

\begin{figure}
	\scalebox{1.0}{\includegraphics[width=0.98\linewidth]{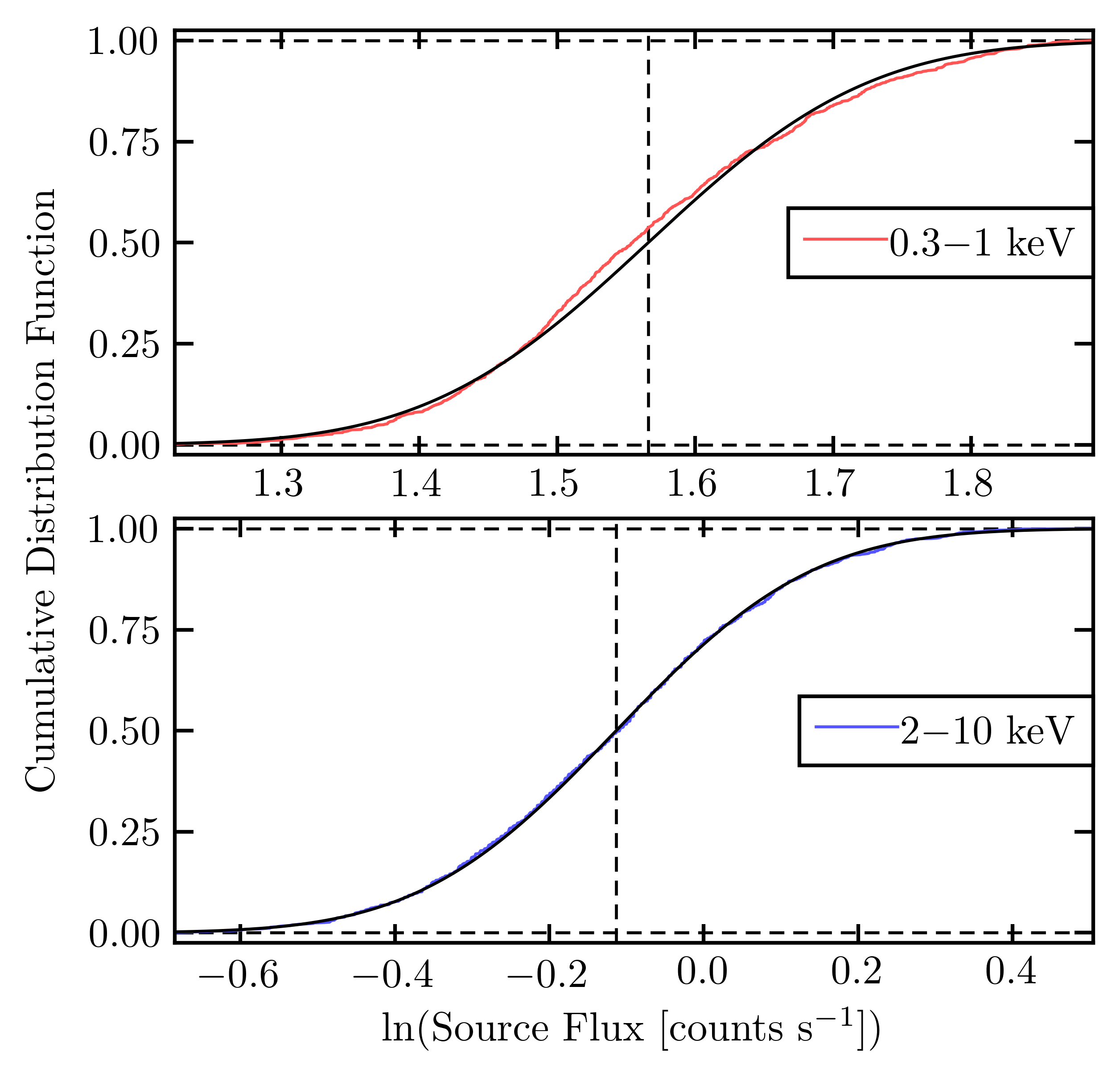}}
	\caption{The natural logarithm of the soft ($0.3-1~\mathrm{keV}$; top panel) and hard ($2-10~\mathrm{keV}$; bottom panel) source flux distributions using light curves binned by $200~\mathrm{s}$. In each panel the black solid line represents the best-fit normal distribution to the data and the vertical black dashed line represents the mean flux.}
	\label{fig:fluxdistros}
\end{figure}

Analysing the flux distributions of a time series can uncover the nature of the variability of that time series. The observed variability in all accreting systems follows a stochastic process which is typically considered \emph{stationary} on the timescale of the observations. All accreting systems are observed to display a log-normal flux distribution (e.g. \citealt{UttleyMcHardyVaughan2005,Alston2019}, and references therein). A significant deviation from a stationary process was first observed in \cite{Alston+2019}, where a non-log-normal flux distribution and a time-dependent power spectral density (PSD) were observed (see also \citealt{Alston2019}). To explore the possibility of non-stationarity in \src we analysed the log-flux distributions of the soft and hard energy bands using $200~\mathrm{s}$ binned light curves, shown as the empirical cumulative distribution functions (ECDFs)\footnote{We use ECDFs here rather than histograms to avoid arbitrarily binning the data in the flux domain.} in Figure \ref{fig:fluxdistros}. The log-flux distributions were each fit with a normal distribution, and the normality of both bands were evaluated by performing Kolmogorov-Smirnov (KS) and Anderson-Darling (AD) \citep{AndersonDarling1952} tests. We found that for the soft band $p_{\mathrm{KS}} = 0.027$ and $p_{\mathrm{AD}} = 3.6\times10^{-7}$ whereas for the hard band $p_{\mathrm{KS}} = 0.99$ and $p_{\mathrm{AD}} = 0.89$. These results may suggest that variability in the soft band is due to physical changes in the system while hard band variability is explained by stochastic variations in the emission process. This interpretation is complicated by the unknown impact of the warm absorption in this source, which significantly affects the soft band below $\sim2~\mathrm{keV}$, on the soft-band flux distribution. Furthermore, it is possible that the $\sim210~\mathrm{ks}$ total observing time is insufficient in providing enough data to adequately sample the flux distributions.

\begin{table*}
	\begin{center}
		\caption{The best-fit models to the hardness-flux and flux-flux plots using $400~\mathrm{s}$ binned light curves with the soft band ($S$) as $0.3-1~\mathrm{keV}$ and the hard band ($H$) as $2-10~\mathrm{keV}$. The columns are: (1) observation abbreviation, (2) best-fit linear model to hardness-flux plot using $F_{\mathrm{C}}$ as the combined flux in the soft and hard bands, (3) the $\chi^2$ value of the given hardness-flux equation, (4) best-fit linear and power-law models to the flux-flux plot using $F_{\mathrm{S}}$ as the soft band flux and $F_{\mathrm{H}}$ as the hard band flux, and (5) the $\chi^2$ value of the given flux-flux equation for the indicated degrees of freedom, $dof$. Errors are not displayed due to the very poor fit qualities, however we note that all parameters have relative errors of $\sim5~\mathrm{per~cent}$ other than the hardness-flux slope estimates which have relative errors of $\sim50~\mathrm{per~cent}$.}
		\begin{tabular}{ccccc}
			\hline
			(1) & (2) & (3) & (4) & (5) \\
			Epoch & Hardness-Flux & Fit Quality & Flux-Flux & Fit Quality \\
			& Equation & $\chi^{2}/dof$ & Equation & $\chi^{2}/dof$ \\
			\hline
			2004.101 & $HR = 0.022F_{\mathrm{C}} -0.77$ & $111/50$ & $F_{\mathrm{S}} = 1.51F_{\mathrm{H}} + 2.83$ & $258/50$ \\
			& & & $F_{\mathrm{S}} = 4.34F_{\mathrm{H}}^{0.32}$ & $256/50$ \\
			&&&&\\
			2004.201 & $HR = 0.021F_{\mathrm{C}} -0.78$ & $125/49$ & $F_{\mathrm{S}} = 1.44F_{\mathrm{H}} + 3.60$ & $266/49$ \\
			& & & $F_{\mathrm{S}} = 5.05F_{\mathrm{H}}^{0.30}$ & $264/49$ \\
			&&&&\\
			2014.101 & $HR = 0.022F_{\mathrm{C}} -0.80$ & $458/198$ & $F_{\mathrm{S}} = 1.17F_{\mathrm{H}} + 3.47$ & $746/198$ \\
			& & & $F_{\mathrm{S}} = 4.63F_{\mathrm{H}}^{0.22}$ & $753/198$ \\
			&&&&\\
			2014.301a & $HR = 0.022F_{\mathrm{C}} -0.86$ & $363/98$ & $F_{\mathrm{S}} = 2.26F_{\mathrm{H}} + 3.32$ & $889/98$ \\
			& & & $F_{\mathrm{S}} = 5.61F_{\mathrm{H}}^{0.38}$ & $865/98$ \\
			&&&&\\
			2014.301b & $HR = 0.010F_{\mathrm{C}} -0.75$ & $193/98$ & $F_{\mathrm{S}} = 2.36F_{\mathrm{H}} + 2.86$ & $570/98$ \\
			& & & $F_{\mathrm{S}} = 5.23F_{\mathrm{H}}^{0.45}$ & $568/98$ \\
			\hline
			\label{tab:lcfits}
		\end{tabular}
	\end{center}
\end{table*}

\begin{figure*}
	\scalebox{1.0}{\includegraphics[width=0.98\linewidth]{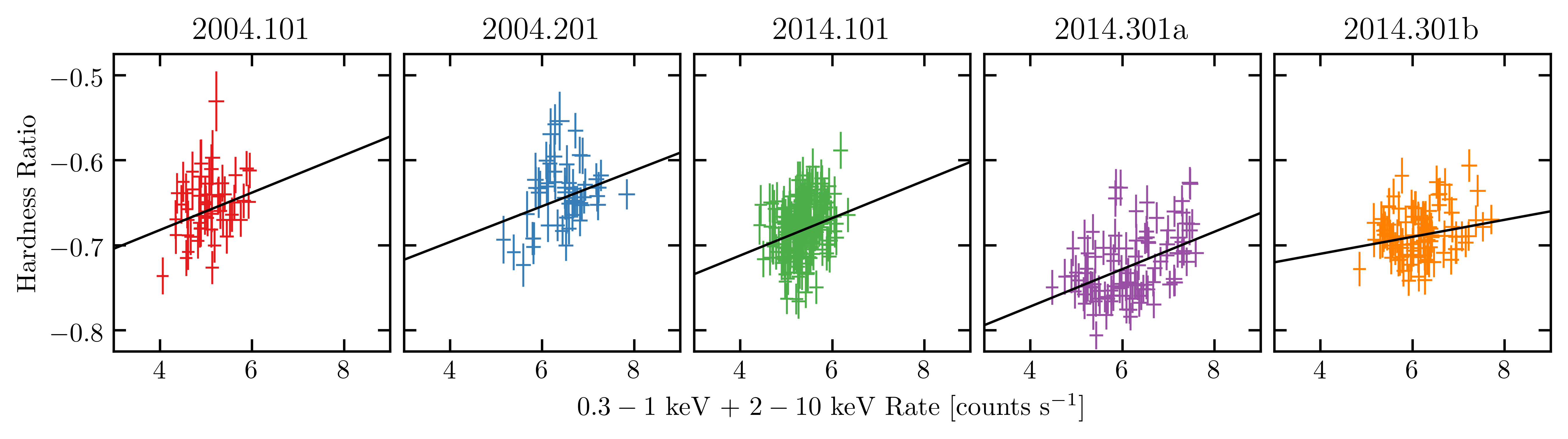}}
	\caption{Hardness-flux plot using $0.3-1~\mathrm{keV}$ and $2-10~\mathrm{keV}$ light curves binned by $400~\mathrm{s}$ as the soft and hard light curves, respectively. In each panel the black solid line displays the best-fit model to the hardness-flux relation, with corresponding equations given in Table \ref{tab:lcfits}.}
	\label{fig:hardnessVflux}
\end{figure*}

Many RQ-NLS1s have been shown to enter a softer spectral state when they are at their brightest (i.e. `softer-when-brighter' trend). To check for such a trend here we constructed a series of hardness ratio versus flux (hereafter simply hardness-flux) plots, shown in Figure \ref{fig:hardnessVflux}. We fit each epoch with its corresponding mean hardness ratio as well as with a linear model with free to vary slope and intercept, finding that the linear model provided a significantly better fit ($>99~\mathrm{per~cent}$ confidence level) in every epoch. The best-fit models are given as column (2) of Table \ref{tab:lcfits}, where in column (3) it is clear that none of the fits provide statistically good descriptions of the data with $\chi^2/dof \gtrsim 2$ in all epochs. Despite the poor fit qualities we note that the `softer-when-brighter' trend is not observed here, in fact the opposite is true: increased flux levels correspond to harder spectra. Furthermore, 2014.301b exhibits a slope that is significantly different from the other epochs, which all share a common slope value of $\sim0.022$. This may suggest that a significant physical change in the central region may have taken place during the 2014.301b epoch. 

\begin{figure*}
	\scalebox{1.0}{\includegraphics[width=0.98\linewidth]{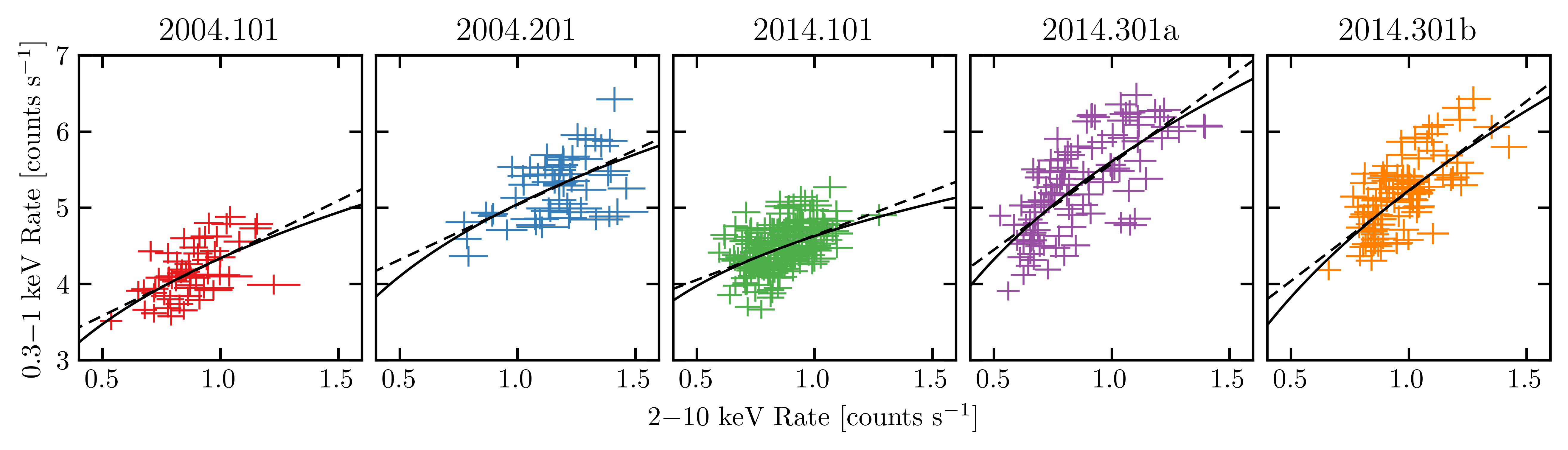}}
	\caption{Flux-flux plots using $0.3-1~\mathrm{keV}$ and $2-10~\mathrm{keV}$ light curves binned by $400~\mathrm{s}$ as the soft and hard light curves, respectively. In each panel the black dashed line displays the best-fit linear model while the black solid line displays the best-fit power-law model, with corresponding equations given in Table \ref{tab:lcfits}.}
	\label{fig:fluxflux}
\end{figure*}

Flux-flux plots have been used as another model-independent timing tool capable of characterising the nature of spectral variability. By plotting the soft versus hard flux we can explore the relationship between these two energy bands directly, as shown in Figure \ref{fig:fluxflux}. We note that in order to avoid arbitrarily binning in the flux-flux plane we only bin the light curves in the time domain. We fit each epoch with both linear (free to vary slope and intercept) and power-law (free to vary normalisation and index) models in order to distinguish the variability as either variations of the hard continuum in the presence of an additional constant component or changes in the hard continuum photon index, respectively \citep{Taylor+2003}. The best-fit linear and power-law models are given as column (4) in Table \ref{tab:lcfits}, where in column (5) it is clear that again none of the fits provide statistically good descriptions of the data with $\chi^2/dof \gtrsim 3$ in all epochs for both models. While a power-law fit is found to the better description overall, suggesting that the spectra exhibit varying photon index values throughout the observation epochs thus providing a straightforward explanation for the hardness ratio variability, we note that the poor fit qualities obtained indicate that the variability observed is due to more than simply a spectral power-law normalisation change and / or pivot. We note that the 2014.301 epochs are much steeper in both model descriptions than the previous three epochs. A power-law plus constant fit (to account for a changing photon index in the presence of an additional constant component) was attempted and found to provide a marginally better fit to the data, though the parameters could not be constrained in any meaningful way and are thus not presented here.

\begin{figure}
	\scalebox{1.0}{\includegraphics[width=0.98\linewidth]{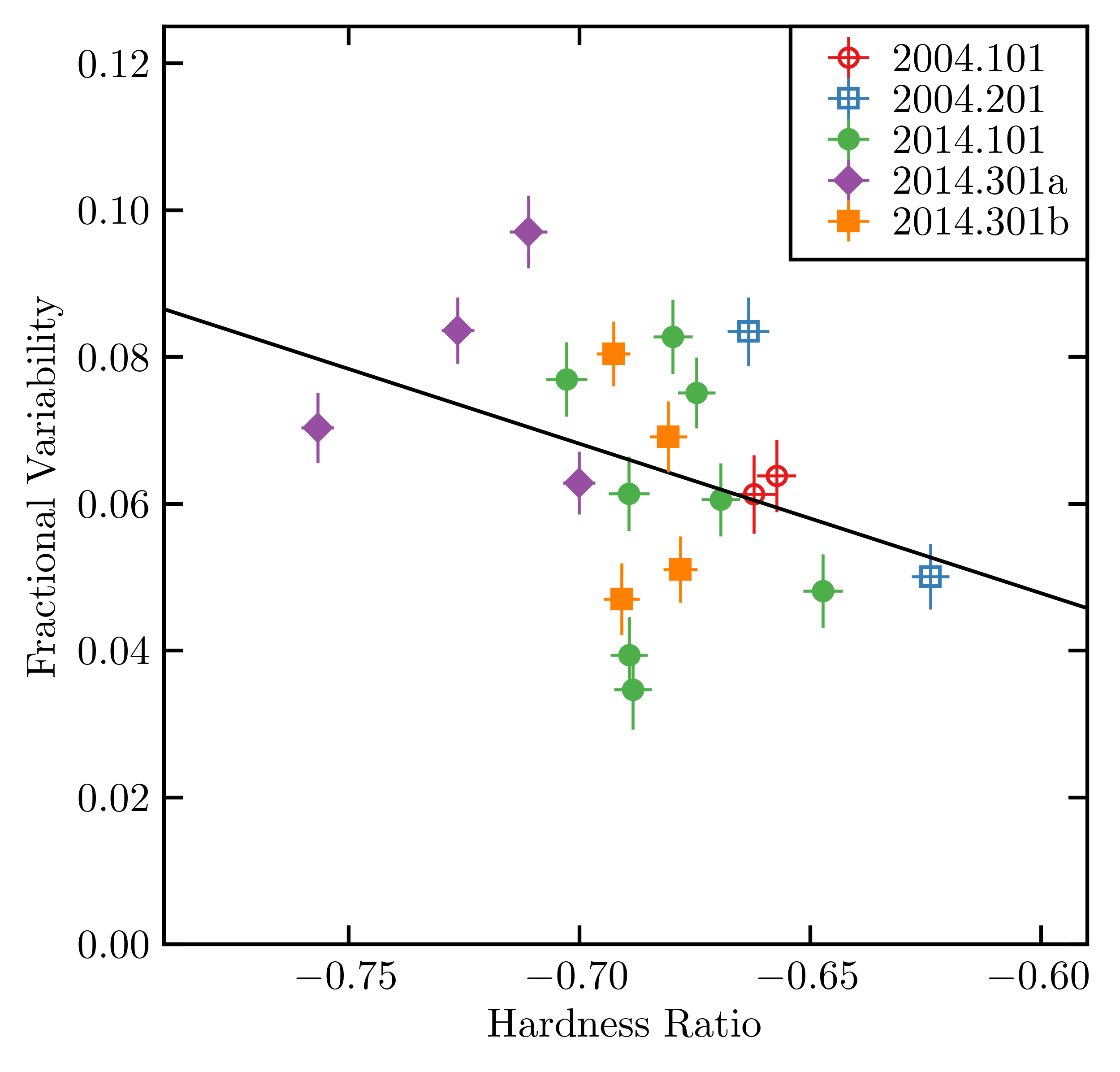}}
	\caption{The mean fractional variability versus mean hardness ratio of $10~\mathrm{ks}$ light curve segments using $400~\mathrm{s}$ binned light curves for each epoch. The best-fit linear model $F_{\mathrm{var}} = \left[-0.2\pm0.1\right]HR + \left[-0.07\pm0.09\right]$ is shown as the solid black line.}
	\label{fig:fvarVShardness}
\end{figure}

In Table \ref{tab:lcprops} we note that the fractional variability increases during the 2014.301a soft state. To explore this further we split each observation into $10~\mathrm{ks}$ segments, binning each by $400~\mathrm{s}$, and computed the fractional variability of the broad band as well as the hardness ratio between the soft and hard bands for each epoch. The results are shown in Figure \ref{fig:fvarVShardness}, where the best-fit linear model was found to be $F_{\mathrm{var}} = \left[-0.2\pm0.1\right]HR + \left[-0.07\pm0.09\right]$ yielding a fit quality of $\chi^2/dof = 272/2$. The fit reveals a trend of decreased variability during harder spectral states, though the fit quality is very poor.

\begin{figure}
	\scalebox{1.0}{\includegraphics[width=0.98\linewidth]{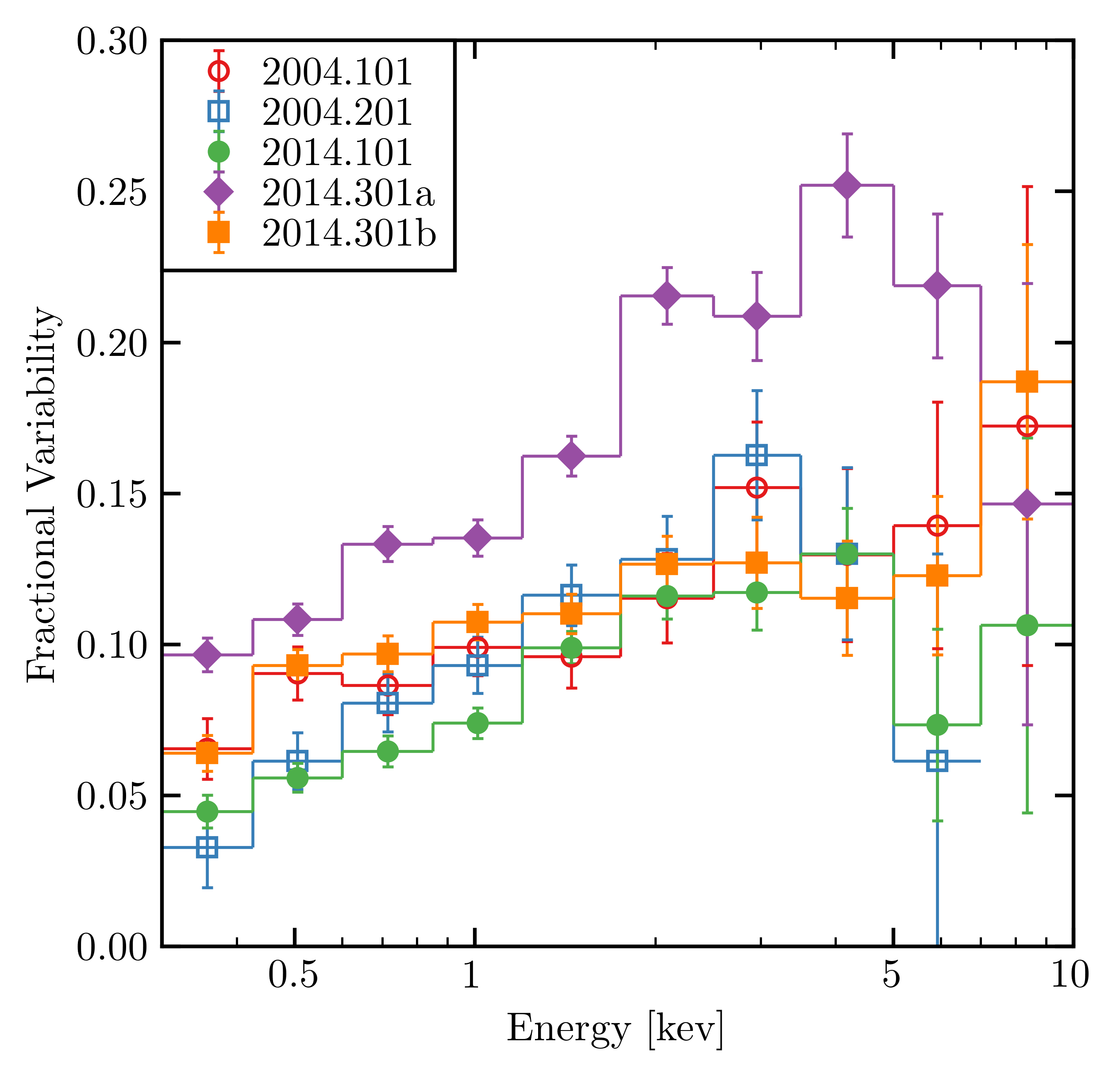}}
	\caption{The fractional variability versus energy spectra using light curves binned by $400~\mathrm{s}$.}
	\label{fig:fvarall}
\end{figure}

Finally, noting the increased variability in the hard band compared to the soft band, we computed fractional variability spectra for each of the five epochs in various energy bins (approximately evenly spaced in energy, combining high-energy bins to reduce error size), shown in Figure \ref{fig:fvarall}. Immediately notable is the consistent shape of the spectra, which all display a trend of increasing variability with energy. Other RQ-NLS1s typically exhibit fractional variability spectra that appear peaked at low energies or are relatively flat across all energies (e.g. \citealt{MarkowitzEdelsonVaughan2003}) which have been interpreted as evidence of a dominant disc component or a dominant power-law changing only in normalisation, respectively. While four of the epochs exhibit similar levels of variability across the entire energy range, 2014.301a displays significantly higher variability at all energies while also more steeply increasing in variability up to $\sim5~\mathrm{keV}$, where the level drops to more like those at other epochs. 

To ensure that differing light curve lengths between the epochs did not produce the observed differences in the fractional variability spectra, we computed the spectra based on $20~\mathrm{ks}$ segments for each epoch. Comparing the mean spectrum of that segmented analysis with the mean spectrum of the presented analysis in Figure \ref{fig:fvarall} found them to be completely consistent.

\subsection{Frequency-space products}
The results so far have been useful to reveal general properties of the various observation epochs. In order to probe the temporal variability of the data in a more detailed manner we must move into frequency based analysis methods. To do this we followed the various prescriptions outlined in \cite{Uttley+2014}, and we summarize the steps here. In this section we focus solely on the two 2014 observations as they are free from significant background flaring and allow for the analysis of the low-frequency regime.

\subsubsection{Combined analysis}

\begin{figure}
	\scalebox{1.0}{\includegraphics[width=\linewidth]{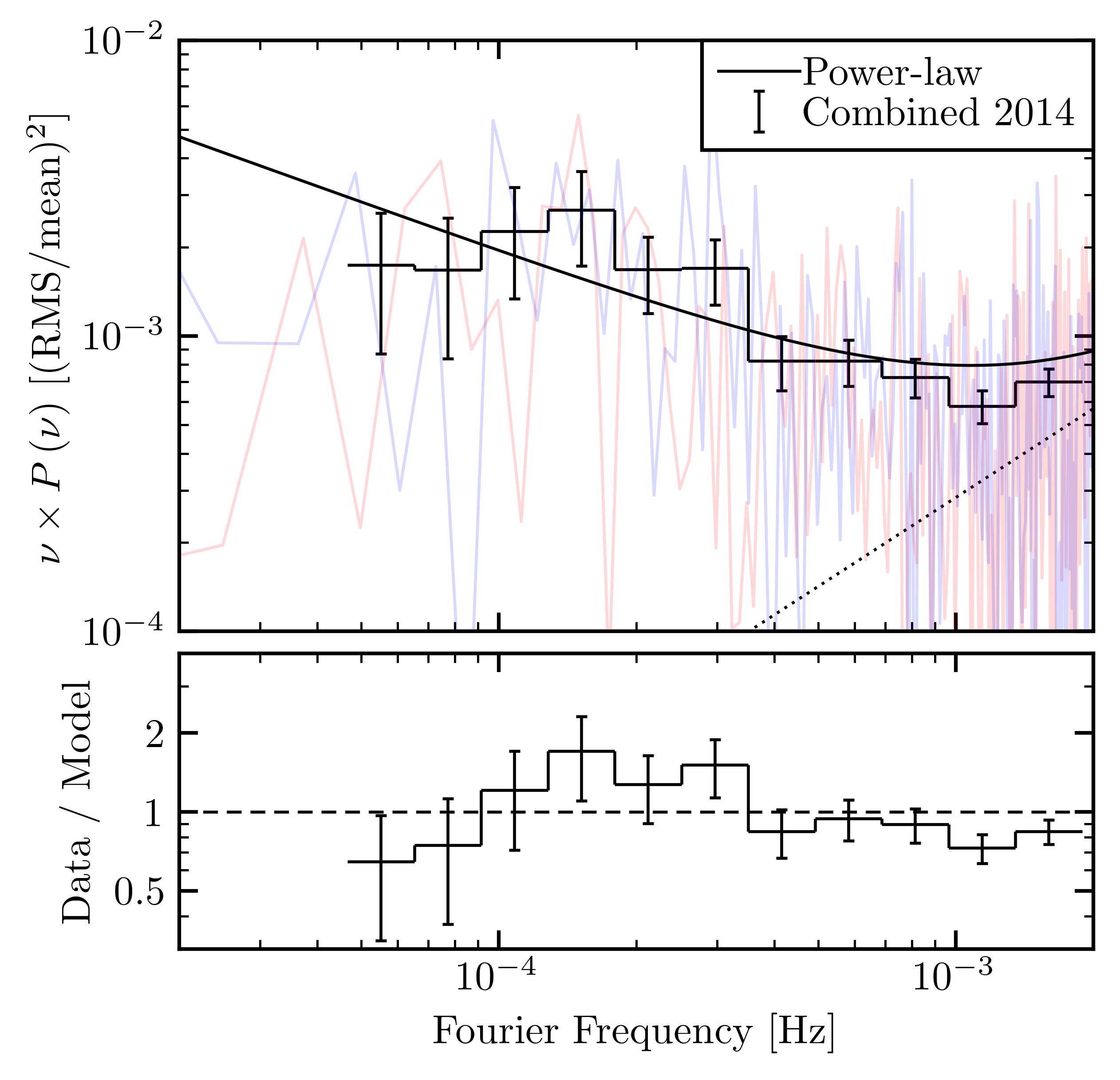}}
	\caption{\textit{Top panel} $-$ The broad band ($0.3-10~\mathrm{keV}$) power spectral density (PSD) using $200~\mathrm{s}$ binned light curves for the combined 2014 epochs, with the raw PSDs for the 2014.101 (red) and 2014.301 (blue) epochs also shown. The Poisson noise level is shown as the dotted line. The solid line displays the best-fit power-law model to the raw PSDs of the 2014 epochs. \textit{Bottom panel} $-$ Ratio residuals when dividing the PSD by the best-fit power-law to the 2014 data.}
	\label{fig:psd}
\end{figure}

We first computed the $0.3-10~\mathrm{keV}$ PSD for the combined 2014 observations. For each epoch, we generated a broad band ($0.3-10~\mathrm{keV}$) light curve binned by $200~\mathrm{s}$. We then performed a discrete Fourier transform (DFT) on each broad band light curve, allowing for a subsequent calculation of the PSD (i.e. RMS normalized square of the periodogram) in each observation. We then combined the two PSDs and binned the result in frequency-space using a geometric binning factor of $1.4$, which is shown in Figure \ref{fig:psd}. We excluded frequencies lower than $3/T~\mathrm{Hz}$, where $T$ is the light curve length, in order to ensure that any periodic signals are detected at least three times and to avoid the impact of red-noise leakage. 

\begin{table}
	\begin{center}
		\caption{Parameter estimates ($90~\mathrm{per~cent}$ confidence intervals) for simultaneous fitting of raw PSDs using a power-law model $P\left(\nu\right) = N\nu^{-\alpha} + C$. The columns are: (1) fit parameter, (2) mean value from MCMC simulations, and (3) and (4) being the $5~\mathrm{per~cent}$ and $95~\mathrm{per~cent}$ bounds from the MCMC simulations, respectively. A superscript $^f$ denotes a model parameter that was kept fixed to the listed value during fitting.}
		\begin{tabular}{cccc}
			\hline
			(1) & (2) & (3) & (4) \\
			Parameter & Mean & $5~\mathrm{per~cent}$ & $95~\mathrm{per~cent}$ \\
			\hline
			Normalisation, $N$ & $1.1\times10^{-5}$ & $4.1\times10^{-6}$ & $2.7\times10^{-5}$ \\
			Slope, $\alpha$ & $1.56$ & $1.44$ & $1.69$ \\
			Poisson noise, $C$ & $0.264^f$ & & \\
			\hline
			\label{tab:psd-params}
		\end{tabular}
	\end{center}
\end{table}

At frequencies $\gtrsim10^{-3}~\mathrm{Hz}$ the source signal is dominated by the Poisson (i.e. white) noise. The overall shape of the PSD resembles those seen in other Type 1 AGN (e.g. \citealt{GonzalezMartinVaughan2012}) where a possible flattening is observed below $\sim1.5\times10^{-4}~\mathrm{Hz}$. To test the significance of this flattening we simultaneously fit the raw (i.e. not frequency binned) PSDs from both epochs with a power-law and a bending power-law model, using the Whittle statistic to evaluate the model likelihoods. Following the methods in \cite{Vaughan2010} we performed the likelihood ratio test (LRT) to compare the two models, finding that the bending power-law model provides $T_{\mathrm{LRT}}^{\mathrm{obs}} = 3.45$ when compared to the power-law model, finding a break frequency in the PSD of $\nu_{\mathrm{break}} \approx 2\times10^{-5}~\mathrm{Hz}$. However, fitting with the bending power-law model did not constrain the low-frequency slope well and thus it was frozen at $\alpha_{\mathrm{L}} = -0.465$, which itself was found to be the best-fit value for this parameter when sampling values between $-1$ and $0$. 

In order to evaluate the significance of the fit improvement offered by the bending power-law model we ran Markov Chain Monte Carlo (MCMC) simulations ($5$ chains of $50,000$ iterations) to draw from the posterior parameter distributions of the power-law fit, which were subsequently used to simulate posterior predictive PSDs. These simulated PSDs were then fit by both the power-law and bending power-law models in order to simulate a distribution of $T_{\mathrm{LRT}}^{\mathrm{sim}}$. Simulating $N=2500$ PSDs from the posterior distribution of power-law parameters returned a posterior predictive $p$-value of $p=0.063$ when comparing $T_{\mathrm{LRT}}^{\mathrm{obs}}$ to $T_{\mathrm{LRT}}^{\mathrm{sim}}$, which itself follows a $\chi^2_1$ distribution. This indicates a modest fit improvement for the inclusion of a bend when simultaneously fitting the raw PSDs but cannot be concluded as statistically significant. Furthermore, the parameter values for the bending power-law model were unable to be meaningfully constrained and thus we do not discuss this model beyond this point. The $90~\mathrm{per~cent}$ confidence intervals for the power-law parameters are given in Table \ref{tab:psd-params} (the Poisson noise level was held constant at the computed noise level during the fitting procedure).

\begin{figure}
	\scalebox{1.0}{\includegraphics[width=\linewidth]{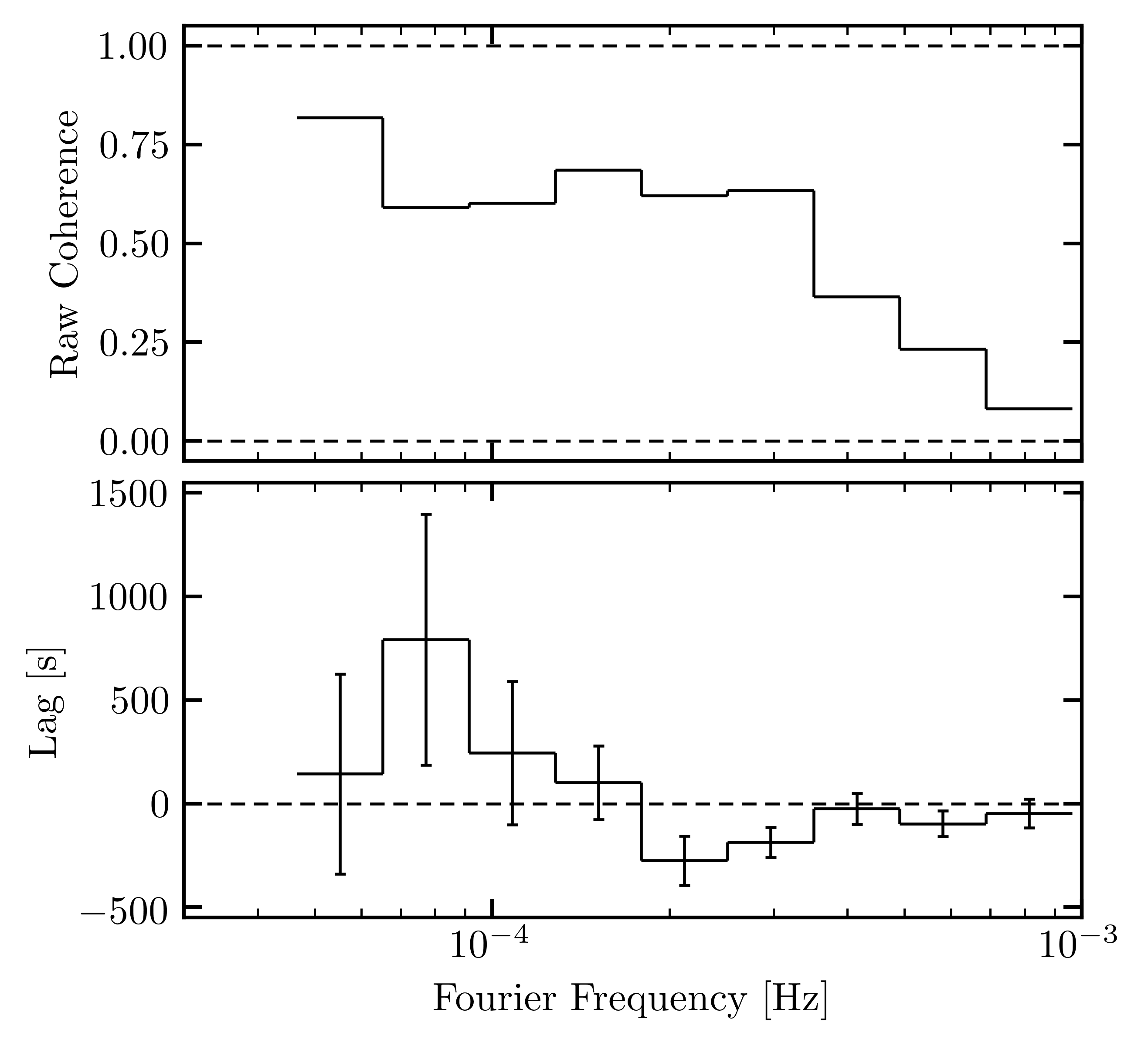}}
	\caption{\textit{Top panel} $-$ The raw coherence of the soft ($0.3-1~\mathrm{keV}$) and hard ($1-4~\mathrm{keV}$) light curves, binned by $200\mathrm{s}$, for the combined 2014 epochs. \textit{Bottom panel} $-$ The lag-frequency spectra computed using the same light curves are shown, where positive lags indicate soft band leading the hard, and negative lags indicate soft band lagging the hard.}
	\label{fig:cohlag}
\end{figure}

The ubiquitous additional constant component found in the flux-flux analysis may be a sign of X-ray reflection off of the accretion disc. This reflector may also be detected via time-lag analysis of two light curves with one being from an energy band dominated by the primary emission and the other being from an energy band dominated by the reflected emission. We therefore computed the combined lag-frequency spectrum to study the time-lag between the soft and hard bands. For each epoch, we generated soft ($0.3 - 1~\mathrm{keV}$) and hard ($1 - 4~\mathrm{keV}$) light curves binning by $200~\mathrm{s}$. We then performed a DFT on each light curve, taking the complex conjugate of the soft light curve DFT and multiplying it by the hard light curve DFT to produce a complex valued product, the cross-spectrum, binning in frequency-space using a geometric binning factor of $1.4$. The argument of this complex valued binned cross-spectrum is the phase $\phi(\nu_i)$, which gives the phase-lag between the soft and hard light curves in the bin at frequency $\nu_i$\footnote{In this work $\nu_i$ is taken to be the bin mid-point frequency.}. From this, we compute the time-lag at frequency $\nu_i$ as $\tau\left(\nu_i\right) = \phi\left(\nu_i\right)/\left(2\pi\nu_i\right)$. To further increase our signal-to-noise we chose to combine the two 2014 observations to produce a single lag-frequency spectrum. This involved combining the cross-spectra of the two light curves from each epoch, and then proceeding with the remaining steps outlined above as normal. The resultant combined lag-frequency spectrum is shown in the bottom panel of Figure \ref{fig:cohlag}, with the corresponding raw coherence plotted in the top panel. As with the PSDs, here we exclude frequencies lower than $3/T~\mathrm{Hz}$.

The combined lag-frequency spectrum exhibits large positive lags in the low-frequency range of $\left(4.66- 12.8\right)\times10^{-5}~\mathrm{Hz}$ and negative lags in the high-frequency range of $\left(1.79 - 3.51\right)\times10^{-4}~\mathrm{Hz}$. We are unable to explore beyond $10^{-3}~\mathrm{Hz}$ due to very low coherence at frequencies greater than this. Notable is the fact that at no point is the coherence $>0.80$, which is at odds with many other AGN that have had their lag-frequency spectra evaluated finding coherence values of $\sim1$ over a broad frequency range. High levels of coherence indicate a simple linear transform between the low- and high-energy input light curves, which may be expected in a strictly reverberation scenario wherein the light curve variability is preserved. The level of incoherence observed here thus indicates a non-linearity in the transform, which may simply be due to Poisson noise. However, given the observed differences in flux distribution between the low- and high-energy regimes the low coherence here may be an independent piece of evidence that suggests physical differences in the emission processes corresponding to these different energy bands. The negative lags present in the lag-frequency spectrum indicate that hard band variations take place first and are `echoed' by the soft band some time later (i.e. the lag value). With our selected soft and hard band energy range this therefore provides evidence of the reverberation of coronal emission off of the accretion disc. 

\begin{figure}
	\scalebox{1.0}{\includegraphics[width=\linewidth]{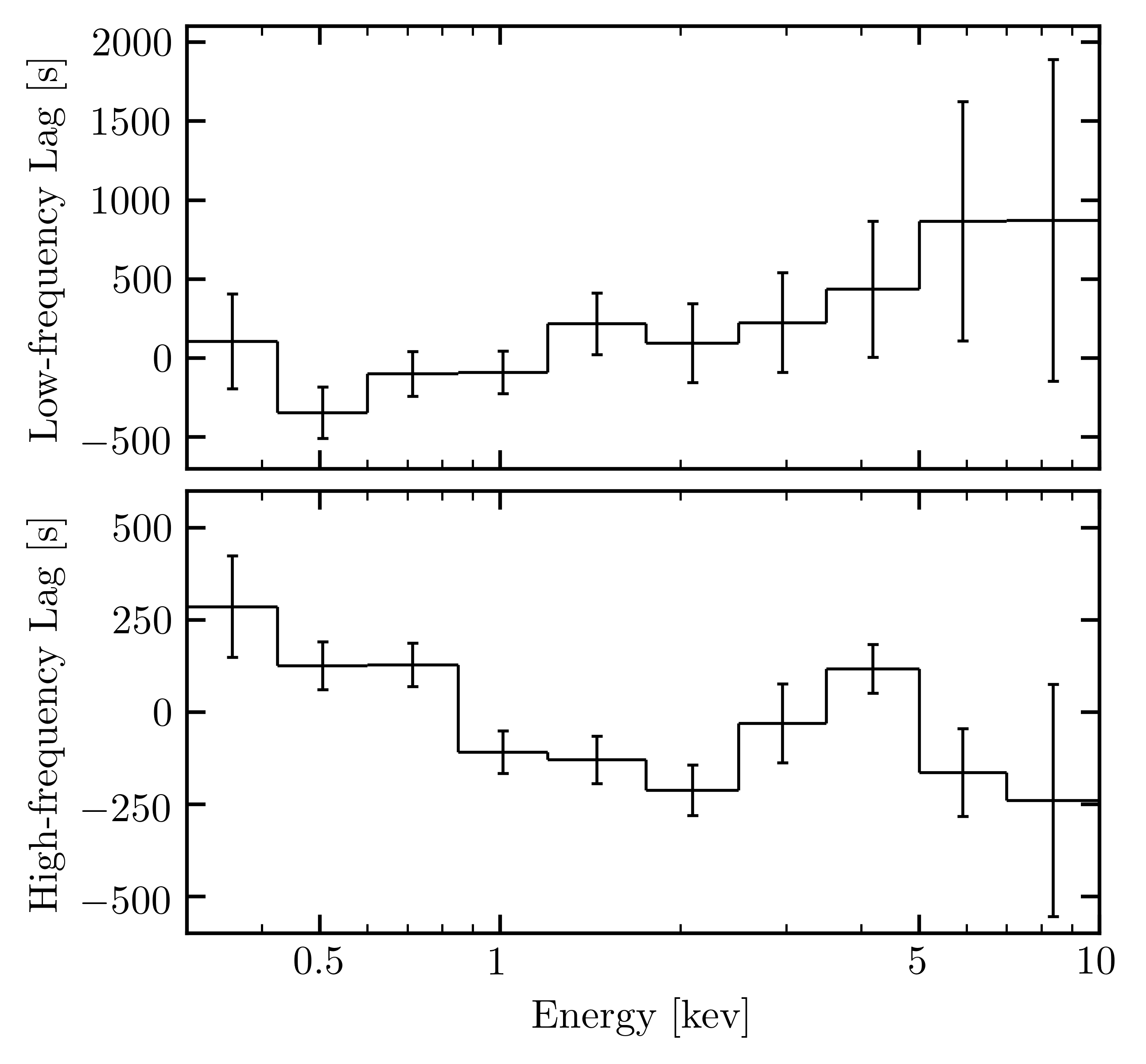}}
	\caption{The combined low-frequency $\left(4.66 - 12.8\right)\times10^{-5}~\mathrm{Hz}$ (top panel) and high-frequency $\left(1.72 - 3.51\right)\times10^{-4}~\mathrm{Hz}$ (bottom panel) lag-energy spectra using $200~\mathrm{s}$ binned light curves with $0.3-10~\mathrm{keV}$ as the reference band.}
	\label{fig:lag-energy}
\end{figure}

Lag-energy spectra were computed in the defined low- and high-frequency bands. We first generated light curves binned by $200$ s for each observation in energy bands that give an approximately equally sampled log-energy-space. For each energy band we subtracted the band-of-interest light curve (i.e. those bound by the energy bin edges) from the reference broad band light curve ($0.3 - 10~\mathrm{keV}$) and computed the cross-spectrum between the two. We then averaged the cross-spectrum in the desired frequency band over the two 2014 observations. With this single cross-spectrum we then computed the lag between the band-of-interest and reference band in the desired frequency range. The resultant low- and high-frequency lag-energy spectra are shown in Figure \ref{fig:lag-energy}. It is important to keep in mind that in these lag-energy plots the lag value on the $y$-axis is not important, rather, the separation between data points is the quantity of interest. This is due to the zero-point being arbitrary as the lag value depends on the dominant spectral component over the broad band which itself is defined differently at each data point due to the aforementioned band-of-interest subtraction. 

The low-frequency lag-energy spectrum presents a shape attributed to the propagation of mass accretion rate changes in the disc through the corona \citep{Lyubarskii1997,KotovChurazovGilfanov2001,ArevaloUttley2006}. In this scenario the outer, low-energy regions of the corona receive these fluctuations before the inner, high-energy regions, thus explaining the observed approximately log-linear relation. Our defined low-frequency band does include two lag values consistent with zero, however, when restricting the analysis to the single bin with a strictly positive lag we retrieved the same lag-energy spectral shape with larger errors. 

The high-frequency lag-energy spectrum presents a shape attributed to X-ray reflection, where the reflected emission (i.e. soft excess at $<2~\mathrm{keV}$ and relativistically broadened Fe K at $4-7~\mathrm{keV}$) changes after the coronal emission (i.e power-law at $2-4~\mathrm{keV}$), hence explaining the large lags at $<1~\mathrm{keV}$ and $3-5~\mathrm{keV}$ and minimal lags at $\sim2~\mathrm{keV}$. Notable is the fact that $5-7~\mathrm{keV}$, which contains the Fe K$\alpha$ emission complex, does not show an increased lag relative to $3.5-5~\mathrm{keV}$, rather it is significantly decreased. This perhaps suggests that we are not sampling long enough time-scales to observe reverberation of the Fe K$\alpha$ emission. Indeed reducing our high-frequency range to lower frequencies finds a corresponding slight increase in the $5-7~\mathrm{keV}$ lag, though a peak in this energy band is never found, thus we present only the wide frequency-band results. This peculiar aspect of the high-frequency lag-energy spectrum is also observed in IRAS 13224$-$3809 \citep{Kara+2016}. Given the interpretation of the high-frequency lag-energy spectrum we are therefore motivated to test a reflection model in our spectral analysis based on this model-independent timing analysis result.

We note that our combined-epoch lag results are consistent with those presented in \cite{Kara+2016}, wherein the reverberation lag in this source was first reported.

\begin{figure}
    \scalebox{1.0}{\includegraphics[width=\linewidth]{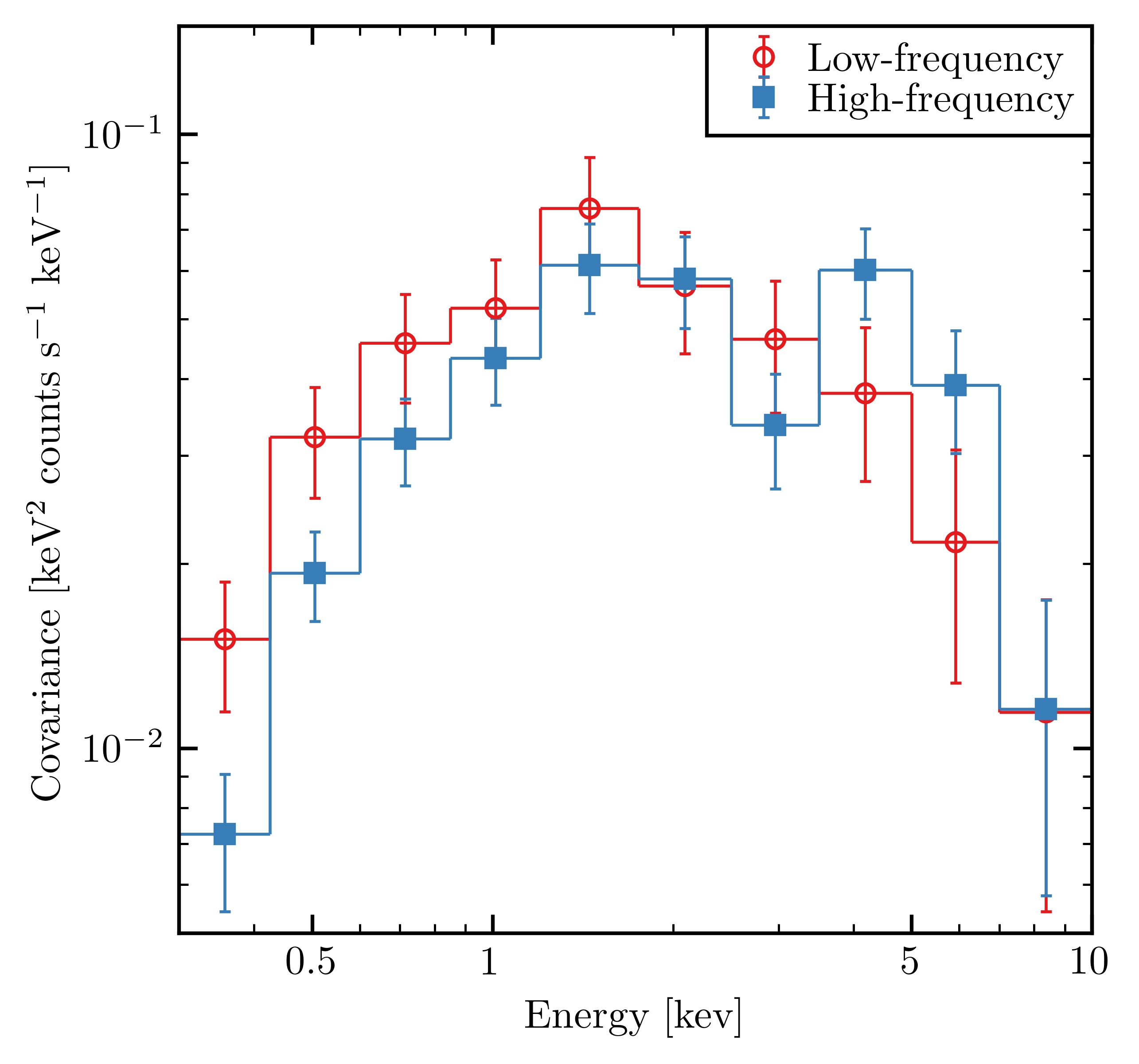}}
	\caption{The combined low-frequency $\left(4.66 - 12.8\right)\times10^{-5}~\mathrm{Hz}$ (red open circles) and high-frequency $\left(1.72 - 3.51\right)\times10^{-4}~\mathrm{Hz}$ (blue filled squares) absolute covariance spectra using $200~\mathrm{s}$ binned light curves with $0.3-10~\mathrm{keV}$ as the reference band. We plot these spectra in $EF_{\mathrm{E}}$ units in order to directly compare the covariance shape to spectral component shapes.}
	\label{fig:covariance}
\end{figure}

Lastly, we calculated the low- and high-frequency (absolute) covariance spectra, which are useful in determining the shape of variable spectral component(s). This was done by using essentially the same base procedure as the lag-energy analysis, making use of the same light curves in the same energy bands with $0.3-10~\mathrm{keV}$ as the reference band. Instead of computing the lags from the cross-spectrum we calculated the covariance as in \cite{Uttley+2014}. The results are shown in Figure \ref{fig:covariance} where both the low- and high-frequency spectra display similar shapes, which suggests a similar varying component across a broad range of time-scales. The shape itself may be interpreted as that resulting from a variable power-law, where the multiple absorption effects impacting this source act to reduce the observed power-law emission in the low-energy regime while at higher energies the natural fall-off of the power-law produces a similar effect in the covariance spectra. We note tentatively here that the low-frequency spectrum appears peaked toward lower energies than the high-frequency spectrum, suggesting a softer power-law varying on longer timescales than a secondary harder power-law that varies more rapidly. We revisit this in Section \ref{sect:spectral} where we explore the broad band spectral properties.

\subsubsection{Time-resolved analysis}

Motivated by the unique behaviour of 2014.301, especially during 2014.301a, we extracted the frequency-space timing products in a time-resolved manner analysing differences between 2014.101, 2014.301a, and 2014.301b. This required splitting each 2014 observation into $40~\mathrm{ks}$ segments, reducing the light curve length by half which therefore means that we were unable to probe the low-frequency band and were restricted to analysing only the high-frequency regime. We combined both segments of 2014.101 in order to reduce clutter in the figures, taking note that all following analyses found that both segments of 2014.101 behaved similarly. Frequency products were binned in frequency-space using a geometric binning factor of $1.4$. We note that the two highest energy bins in the combined analysis were merged here as individually they returned very large errors on the computed quantities.

\begin{figure*}
	\scalebox{1.0}{\includegraphics[width=0.98\linewidth]{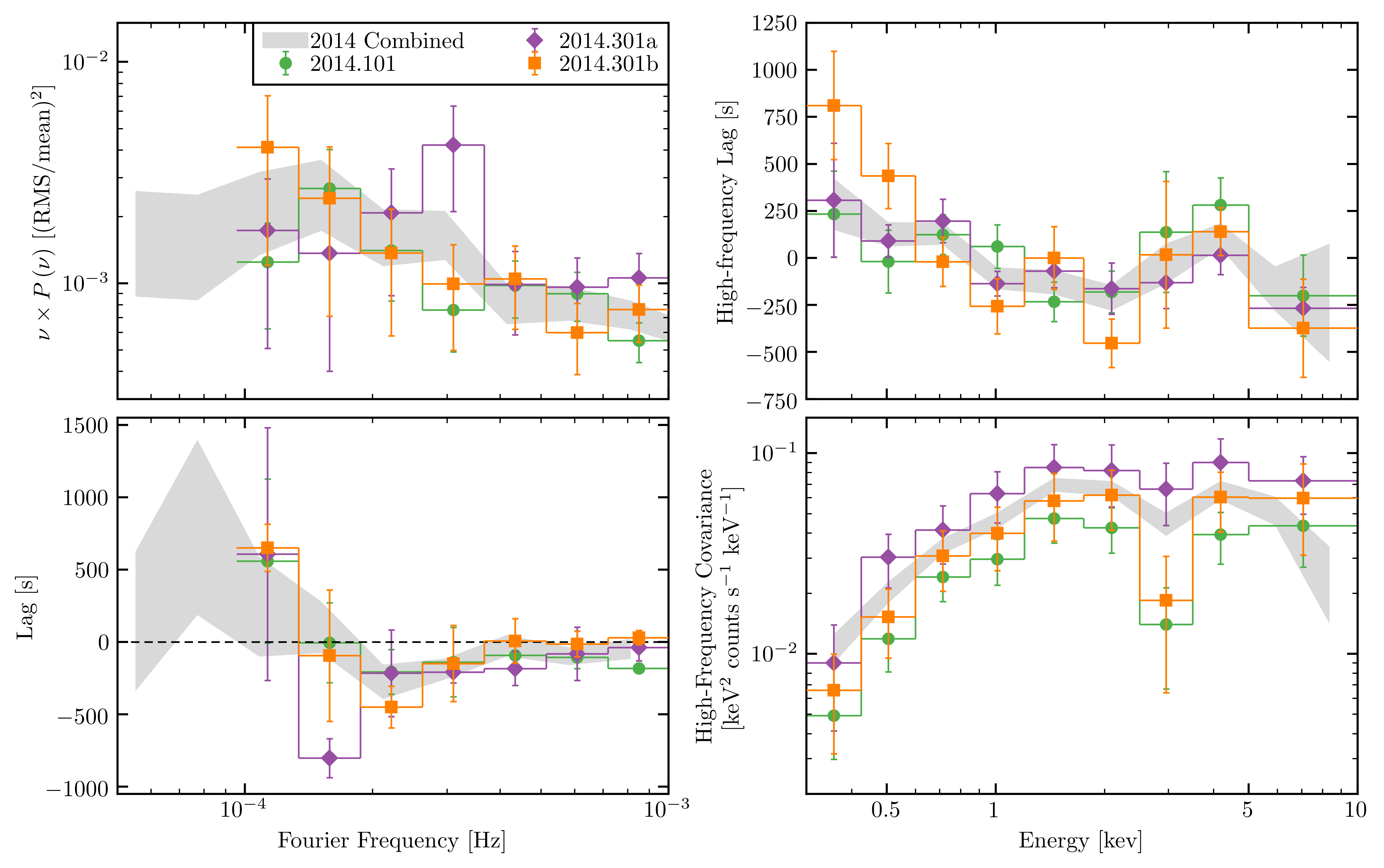}}
	\caption{\textit{Top left} $-$ PSDs for 2014.101 (green circles), 2014.301a (purple diamonds), and 2014.301b (orange squares). \textit{Bottom left} $-$ Lag-frequency spectra for the same epochs in the left panel. \textit{Top right} $-$ High-frequency $\left(1.79 - 3.51\right)\times10^{-4}~\mathrm{Hz}$ lag-energy spectra for same three epochs as other panels. \textit{Bottom right} $-$ Absolute covariance spectra extracted from the same high-frequency band as the lag-energy spectra for the same three epochs as in the other panels, plotted in $EF_{\mathrm{E}}$ units. The grey filled region in each panel displays relevant quantity obtained when combining the 2014 data sets.}
	\label{fig:split-lag-freq}
\end{figure*}

We started by computing broad band $0.3-10~\mathrm{keV}$ PSDs for the three epochs, shown in the top left panel of Figure \ref{fig:split-lag-freq}. The PSD shapes for all epochs are quite similar, with the exception of the bin centred at $\sim3\times10^{-4}~\mathrm{Hz}$, for which 2014.301a displays increased power. Evaluating the significance of this deviation found it to be statistically insignificant when fitting the raw PSDs and comparing parameters, which is perhaps not surprising given that the differences between the PSDs do not exceed the $2\sigma$ level. We also computed the ratio between the 2014.301a epoch and the other two, which did not find the deviations seen in Figure \ref{fig:split-lag-freq} to be statistically significant, again likely owing to the large error bars.

Next we computed the lag-frequency spectra, applying the same methodology described in the previous section, which are shown in the bottom left panel of Figure \ref{fig:split-lag-freq}. All three epochs exhibit negative lags in the $\left(2-4\right)\times10^{-4}~\mathrm{Hz}$ band, which agrees with the combined analysis results. Interestingly, 2014.301a exhibits a larger negative lag at a lower frequency than any other epoch, suggesting a marked change in the behaviour of the source during this epoch, although the change does not exceed the $2\sigma$ level. We note that this observed change in the shape of the lag-frequency spectrum is reminiscent of the results presented by \cite{Alston+2020}, where $\sim2~\mathrm{Ms}$ of data on the RQ-NLS1 IRAS 13224$-$3809 were used to map a moving corona. The results here may therefore suggest a coronal height increase during the 2014.301a epoch. Other RQ-NLS1s such as I Zw 1 \citep{Wilkins+2017} and Mrk 335 \citep{WilkinsGallo2015} have also shown evidence of moving coronae in the context of failed jet launching scenarios. Alternative interpretations for this result are unclear as the frequency at which reverberation lags are detected is thought to be governed by $M_{\mathrm{BH}}$ in systems where X-ray reflection is substantial.

The lag-energy spectra were extracted from the same $\left(1.79 - 3.51\right)\times10^{-4}~\mathrm{Hz}$ band as in the combined analysis, and are shown in the top right panel of Figure \ref{fig:split-lag-freq}. All three epochs exhibit spectra closely resembling the combined result across the entire energy range. We note, however, a slight shape change in the spectral shape over time. In 2014.101 similar lags for energies $\lesssim1~\mathrm{keV}$ and $3.5-5~\mathrm{keV}$ are found. If we interpret the high-frequency lag-energy spectrum as a signature of reverberation this suggests that the emission in these two energy-bands originate at similar distances from the hard X-ray corona, which leads the reverberated emission. In 2014.301b the low-energy lag is significantly increased relative to the high-energy lag. Indeed by taking $\Delta\tau = \tau_{0.3-0.85} - \tau_{2.5-5}$ we find that $\Delta\tau_{\mathrm{2014.101}} = 0 \pm 203$, $\Delta\tau_{\mathrm{2014.301a}} = 257 \pm 141$, and $\Delta\tau_{\mathrm{2014.301b}} = 330 \pm 120$ indicating a significant change in the time delay between these two energy bands.

Absolute covariance spectra were also constructed for each epoch using the same high-frequency range as the lag-energy, with the results shown in the bottom right panel of Figure \ref{fig:split-lag-freq}. All three epochs exhibit the same shape as the combined high-frequency result, indicating the same spectral component is varying in each epoch but with different levels of variability, where in fact the normalisations of the spectra mimic the fraction variability results in column (3) of Table \ref{tab:lcprops}. We note that the only source of deviation in shape between the epochs is in the bin centred at $\sim3~\mathrm{keV}$, which exhibits increased levels of variability in 2014.301a. As aforementioned the shape of these covariance spectra suggest a straight-forward explanation to the hardness ratio dip observed in 2014.301a as the simple variability of the power-law component.

\section{Spectral Analysis}
\label{sect:spectral}
All spectral fitting was done using \textsc{xspec} version 12.10.0c \citep{Arnaud1996} and all spectra were grouped using the \textsc{heasoft} task \textsc{ftgrouppha} with the \textsc{grouptype=opt} flag to bin the spectra as described in \cite{KaastraBleeker2016}. We evaluate the goodness-of-fit of our models by following the method of \cite{Kaastra2017} making use of Cash ($C$) statistics \citep{Cash1979} to evaluate model likelihoods. In order to make proper use of $C$-statistics we are required to use the total (i.e. source plus background) spectrum in each epoch, which differs from the commonly used background-subtracted spectral modelling approach. To do this we simultaneously modelled the total and background spectra\footnote{We followed the guide to source and background modelling at \href{https://heasarc.gsfc.nasa.gov/xanadu/xspec/manual/node40.html}{https://heasarc.gsfc.nasa.gov/xanadu/xspec/manual/node40.html}}, which allowed for the contribution of the background emission to be `modelled-out' of the total spectrum. The fits presented in this section were therefore performed by fitting the five total and five background spectra, linking the total and background spectra for each epoch.

\subsection{Modelling the broad band continuum}
\begin{figure*}
	\scalebox{1.0}{\includegraphics[width=\linewidth]{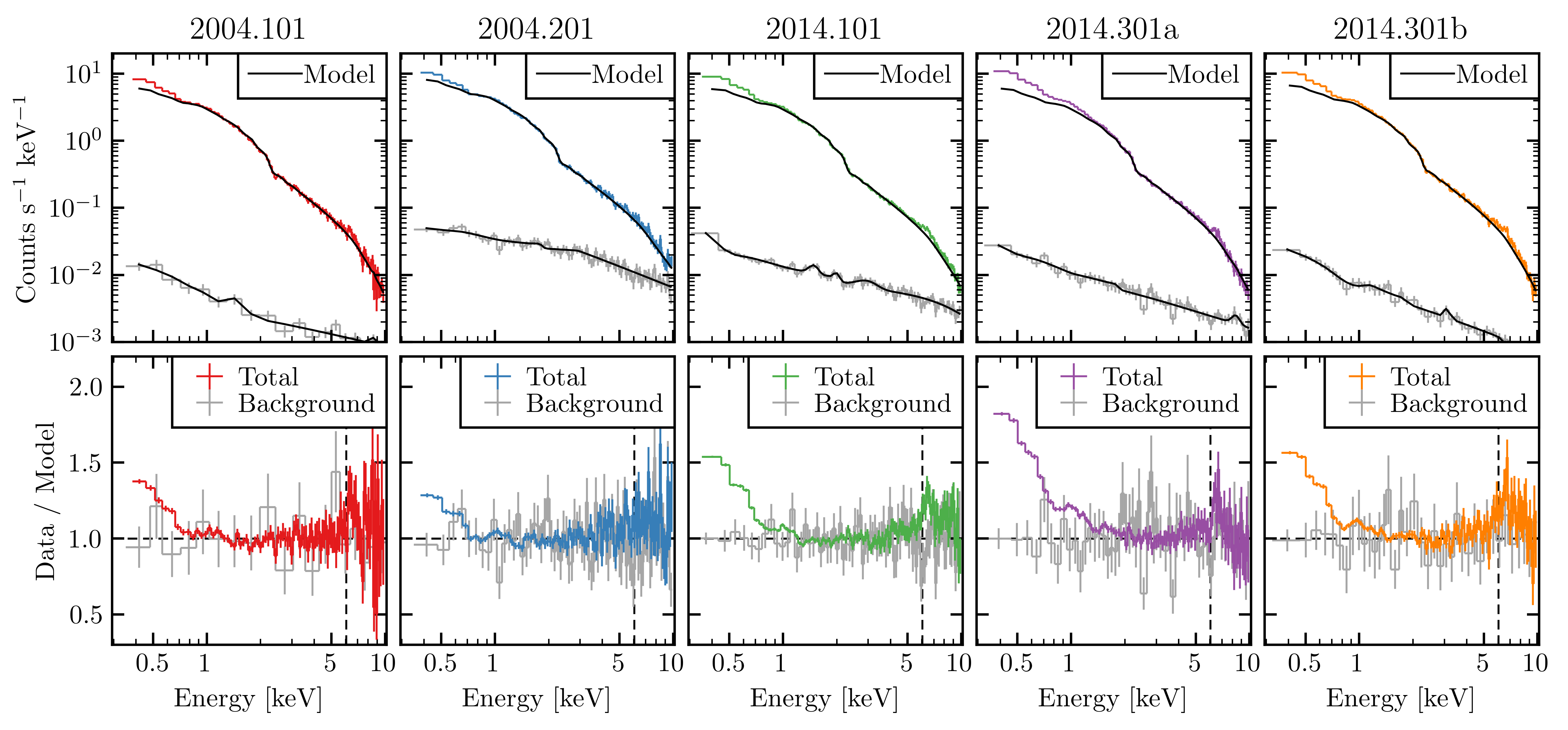}}
	\caption{\textit{Top row} $-$ Coloured (grey) curves are the total (background) spectrum in each epoch with the solid black lines denoting the absorbed power-law ($\Gamma = 2.35\pm0.03$) fit over $2-4~\mathrm{keV}$ (full broad band background model, see text for details). \textit{Bottom row} $-$ Coloured (grey) curves denote the ratio residuals of the total (background) spectrum in each epoch when divided by the model displayed in the top row of the corresponding column. Vertical lines in each bottom panel denote the observed frame Fe K$\alpha$ emission line energy, which has a rest frame energy of $6.4~\mathrm{keV}$.}
	\label{fig:plldra-po24}
\end{figure*}

When performing our spectral fits we first fit the background spectra, starting off with each being fit by a broken power-law. This was found to be insufficient in describing the overall shape of the background spectra, requiring the addition of known instrumental emission and absorption features of the EPIC pn detector at corresponding energies using the \textsc{gauss} model. The background extraction region in each epoch was not found to contain any other X-ray sources and thus we do not add any other components to account for the background emission. After accounting for the continuum shape and instrumental features the fit quality to the background was sufficiently good (i.e. additional components did not improve the fit in a statistically significant way) such that we froze all background model parameters to the fitted values found in this step for the remainder of the spectral modelling. We do not present the background model parameters here, though the fits are presented in Figure \ref{fig:plldra-po24} alongside the total spectrum fits.

After an adequate background model was obtained for each epoch the total spectra were then fit by including the background components in addition to the desired source model components. We initially modelled the continuum by fitting the $2-4~\mathrm{keV}$ band with an absorbed power-law using the \textsc{cutoffpl} model, accounting for Galactic absorption along our line of sight with \textsc{tbabs}. However, this source is known to have significant warm absorption \citep{Leighly+1997,Longinotti+2015,Sanfrutos+2018} that is accounted for here by using the \textsc{xspec} implementation of the \textsc{spex} \citep{KaastraMeweNieuwenhuijzen1996} model \textsc{xabs} \citep{Steenbrugge+2003} provided by Parker et al. (in prep.). We inserted the warm absorbers in order of significance according to \cite{Sanfrutos+2018}, finding that beyond three components the fit did not improve significantly. We initially kept the warm absorber parameters (i.e. column density, ionisation parameter, turbulent velocity, and outflow/inflow velocity) fixed to the values derived in \cite{Sanfrutos+2018}. Due to the flux variability observed throughout Section \ref{sect:timing} we left each power-law model component to have a free to vary normalisation, while linking the photon index across the epochs and freezing the cut-off energy at $E_{\mathrm{cut}} = 300~\mathrm{keV}$ (all fits were found to be insensitive to this parameter). We found a photon index of $\Gamma = 2.35\pm0.03$ and show the ratio residuals (i.e. data divided by the model residuals) for this fit in Figure \ref{fig:plldra-po24} where we extrapolate across the broad $0.3 - 10~\mathrm{keV}$ band.

All five spectra show evidence of a soft excess below $\sim1.5~\mathrm{keV}$, with the strength of this excess being variable over the observations on long and short time scales. Both 2004 epochs exhibit weaker soft excesses than either of the 2014 epochs, with 2014.301a displaying the strongest soft excess of any data set while 2014.101 and 2014.301b appear intermediate. The excess emission observed around the Fe K emission complex at $6-7~\mathrm{keV}$ appears persistent on long and short time scales, though 2004.201 does not visually reveal its presence due to lower spectral quality.

In the following, we outline the approach and motivation for allowing certain spectral parameters to be free to vary during the fitting procedure. The final model, with parameter values and $90~\mathrm{per~cent}$ confidence intervals, are reported in Table \ref{tab:params-final}. 

Based on the reverberation evidence obtained from Section \ref{sect:timing} and the signatures presented in the residuals (i.e. soft excess and broad Fe K emission) we were motivated to fit our data with an additional reflection component. We used \textsc{relxilld} version 1.2.0 \citep{Garcia+2014} to account for relativistically blurred emission originating from the highly ionised inner accretion disc region, with an initial parameter setup as follows. The emissivity profile of the disc (approximated as a once-broken power-law in \textsc{relixlld}) was set such that the inner index $q_{\mathrm{in}}$ was free to vary while the break radius was frozen at $R_{\mathrm{break}} = 6~r_g$ (i.e. the innermost stable circular orbit (ISCO) of a Schwarzschild black hole), where $r_g = GM/c^2$, and the outer index was frozen at $q_{\mathrm{out}} = 3$. We froze the black hole spin at $a=0.998$, freezing the inner disc radius at $R_{\mathrm{in}} = R_{\mathrm{ISCO},0.998} = 1.235~r_g$ and the outer disc radius at $R_{\mathrm{out}} = 400~r_g$. The iron abundance and disc density were initially frozen at $A_{\mathrm{Fe}} = 1$ and $\log N = 15~\mathrm{cm^{-3}}$, respectively. The remaining \textsc{relxilld} parameters of disc inclination $i$ and ionisation parameter $\log \xi$ were left free to vary. We allowed the flux of the power-law and reflection components to be free to vary between epochs, with all other parameters linked across all epochs, and used the power-law model as the emission source illuminating the accretion disc (i.e. $\Gamma_{\mathrm{\textsc{relxilld}}} = \Gamma_{\mathrm{\textsc{cutoffpl}}}$). The resulting fit returned $C = 1761$ for $dof=893$ and presented significant residuals around the Fe K emission complex and higher energies, while the soft excess was well accounted for. 

The evidence from Section \ref{sect:timing} indicates a significant spectral state change during 2014.301a and thus we allowed $\Gamma$ to be free to vary in this epoch only, returning $\Delta C = -329$ for $1$ additional free parameter. The photon index for 2014.301a at this stage in the fitting procedure is significantly softer than the average. Taking into account the global hardness ratio variations of $\sim20~\mathrm{per~cent}$ we then tested allowing $\Gamma$ to be free to vary in all epochs, returning $\Delta C = -127$ for $3$ additional free parameters. Epoch 2014.301a maintained the softest power-law, with the other epochs displaying similar photon index values. This intermediate model provides a mediocre fit to the data with a relatively simple explanation for the spectral variability.

Accretion disc density has recently been shown to be anti-correlated with $M_{\mathrm{BH}}$, with low-mass systems such as NLS1s more frequently exhibiting $\log N > 15~\mathrm{cm^{-3}}$ \citep{Jiang+2019}. We first tested for the possibility of non-solar iron abundance in our data by allowing $A_{\mathrm{Fe}}$ to be free to vary, finding a best-fit value consistent with a solar abundance while not providing a significant fit improvement. Due to this we kept $A_{\mathrm{Fe}} = 1$ frozen throughout the remainder of the spectral fitting. To test for the presence of a high-density disc we allowed $\log N$ to be free to vary, returning $\Delta C = -130$ for $1$ additional free parameter. The disc density found by this fit is significantly higher than the standard accretion disc density, falling in line with other high-density discs found in NLS1s.

Truncated accretion discs with $R_{\mathrm{in}} > R_{\mathrm{ISCO}}$ have been suggested in RL AGN wherein the inner part of disc becomes unstable and subsequently is ejected along the jet resulting in a radio flare and truncated disc (e.g. \citealt{Lohfink+2013}). Here we tested for inner disc truncation by allowing $R_{\mathrm{in}}$ to be free to vary, returning $\Delta C = -10$ for $1$ additional free parameter. This fit does not find a significantly truncated disc, only being $\sim1~r_g$ beyond the ISCO of a maximally spinning Kerr black hole of, though the improvement is substantial.

We also tested the impact of a free to vary break radius in the once-broken emissivity profile of the disc. This parameter was found to modestly improve the fit quality returning $\Delta C = -5$ for $1$ additional free parameter and produced comparable values for the other free parameters as when it was fixed. We therefore kept the value fixed at $R_{\mathrm{break}} = 6~r_g$ for the remainder of the spectral modelling. Black hole spin was also set to be free to vary, though we found no significant fit improvements in doing so and therefore keep it fixed at $a=0.998$ for the remainder of the spectral modelling.

We have only reported detailed results for the reflection scenario, although several alternative models based on other physical scenarios were also tested for their ability to describe the data. Our best-fit model using the reflection scenario so far was found to have $C/dof = 1165/887$. In RL sources the contribution to the X-ray spectrum by the radio jet may be modelled via a double power-law continuum, wherein the X-ray corona is responsible for producing the softer power-law while the harder power-law is representative of the jet emission. We therefore applied a double power-law model here but found it to be a poor fit with $C/dof = 1475/887$. We also tried a soft-Comptonisation model using \textsc{comptt} \citep{Titarchuk+1994} wherein the soft excess can be explained by an optically thick `warm' corona while the high energy emission is due to the standard hot corona, a scenario that has been successfully applied to RQ sources (e.g. \citealt{Petrucci+2018}). We found, however, that such a model was also substantially worse than the reflection scenario returning a best fit of $C/dof = 1392/889$. We do not report any of the parameters of these alternative models as they were found to be statistically worse fits to the data.

\subsection{Iron emission features}
All of the fits discussed up to this point have had significant excess residuals present in the Fe K emission complex around $6-7~\mathrm{keV}$, despite this region having already been treated by the reflection component in our model. In order to account for the poorly fit feature(s) present in this region we tested two cases: (i) a series of intrinsically narrow emission features to emulate Fe \textsc{i, xvii, xvii} emission features, and (ii) a single broad Fe K emission feature. We found that in case (i) by inserting lines at $6.4$, $6.7$, and $6.97~\mathrm{keV}$ using the \textsc{zgauss} model with narrow widths of $\sigma=10~\mathrm{eV}$ and free to vary normalisation linked across all epochs the fit returned $\Delta C = -84$ for $3$ additional free parameters. Notably, line normalisation increased with line energy, perhaps suggesting that a single broad, asymmetric line profile would be a more fitting description. Nevertheless, we also tested accounting for the potentially narrow emission features with the presence of an additional reflection component originating at much larger distances, emulating outer disc / toroidal emission, via the \textsc{xillver} \citep{GarciaKallman2010} model, though we found no significant fit improvement in doing so. For case (ii) we inserted a broad line with free to vary energy, width, and normalisation that were linked across all epochs. This fit returned $\Delta C = -130$ for $3$ additional free parameters, providing a significantly better description of the Fe K emission region than the three narrow lines and returning a line energy of $E\approx6.8~\mathrm{keV}$ and line width of $\sigma\approx0.5~\mathrm{keV}$. Evaluating the ratio residuals produced by this broad line, however, revealed that the symmetric nature of the Gaussian profile over-estimated the the low-energy side of the line. 

The evidence for an asymmetric line profile, which indicates relativistically broadened line emission, prompted the use of the \textsc{laor} model. We allowed the line energy, emissivity index, and inner radius of the disc to be free to vary, linking the outer disc radius and disc inclination to the corresponding \textsc{relxilld} parameters. Upon fitting the spectra it was found that the emissivity index was not statistically differentiable from $q=3$, and thus this parameter was kept fixed to that value. The resulting fit returned $\Delta C = -130$ for $3$ additional free parameters, thus being equivalent to a single broad Gaussian, with much of the asymmetry in those residuals eliminated. We tested allowing the energy and normalisation of the line to be free to vary between the epochs, but found no significant improvements to the fit when doing so and therefore kept these parameters linked across all epochs. We report the results obtained using the \textsc{laor} model though there was no statistical benefit in using this emission line model over, for example, the \textsc{diskline} model which returned similar parameter values.

With the fit described so far we noted the presence of significant residuals at $\sim1~\mathrm{keV}$ in all of the spectra. The cause for such deviations from the model may have been due to either (i) the presence of excess Fe L emission or (ii) improperly modelled warm absorption due unknown RGS to EPIC pn cross calibration. We first tested case (i) by inserting a second \textsc{laor} component at $\sim1~\mathrm{keV}$ with free to vary energy and normalisation, each linked across all epochs, linking all other parameters to those of the Fe K feature found above, resulting in $\Delta C = -68$ for $2$ additional free parameters. The fit is significantly improved in this energy band with the addition of this Fe L feature, though some significant residuals are still present. Testing case (ii) did not require a second emission feature but instead we allowed the column density and ionisation parameter of each of the three warm absorbers to be free to vary, which resulted in $\Delta C = -132$ for $6$ additional free parameters. This significantly improved the residuals across the entire low-energy band. We considered it possible, however, that a truly present Fe L feature has now been washed out artificially by the free warm absorbers. To test this we combined both cases (i) and (ii) by inserting an Fe L feature using a \textsc{laor} component while simultaneously allowing the warm absorbers to each have free to vary column density and ionisation parameter, returning $\Delta C = -164$ for $8$ additional free parameters. The evidence presented here suggests that Fe L may in fact be present in the spectra, although the warm absorbers produce a significantly greater impact on the fit quality and account for the majority of the deviations from the model.

\subsection{The final fit}

\begin{table*}
	\begin{center}
		\caption{Fit parameters and errors (i.e. $90~\mathrm{per~cent}$ confidence intervals) for the final reflection model. Parameters flagged with a $^f$ were kept frozen to the listed value and those flagged with a $^l$ were kept linked to the indicated parameter throughout the fitting procedure.}
		\begin{tabular}{ccccccc}
			\hline
			Model Component & Parameter & \multicolumn{5}{c}{Fit Value} \\
			& & 2004.101 & 2004.201 & 2014.101 & 2014.301a & 2014.301b \\
			\hline
			\textsc{tbabs} & $N_{\mathrm{H}} / 10^{20}~\mathrm{cm^{-2}}$ & \multicolumn{5}{c}{$2.20^{f}$} \\

			$\times$ & & & & & & \\

			\textsc{xabs1} & $N_{\mathrm{H}} / 10^{20}~\mathrm{cm^{-2}}$ & \multicolumn{5}{c}{$13^{+1}_{-2}$} \\
			& $\log\xi$ & \multicolumn{5}{c}{$1.50\pm0.07$} \\
			& $v_{\mathrm{turb}}/\mathrm{km~s^{-1}}$ & \multicolumn{5}{c}{$60^f$} \\
			& $v_{\mathrm{outflow}}/\mathrm{km~s^{-1}}$ & \multicolumn{5}{c}{$140^f$} \\
			& $z$ & \multicolumn{5}{c}{$0.0604^f$} \\

			$\times$ & & & & & & \\			

			\textsc{xabs2} & $N_{\mathrm{H}} / 10^{20}~\mathrm{cm^{-2}}$ & \multicolumn{5}{c}{$10^{+0.9}_{-1.6}$} \\
			& $\log\xi$ & \multicolumn{5}{c}{$0.9\pm0.1$} \\
			& $v_{\mathrm{turb}}/\mathrm{km~s^{-1}}$ & \multicolumn{5}{c}{$1000^f$} \\			
			& $v_{\mathrm{outflow}}/\mathrm{km~s^{-1}}$ & \multicolumn{5}{c}{$110^f$} \\
			& $z$ & \multicolumn{5}{c}{$0.0604^f$} \\

			$\times$ & & & & & & \\			

			\textsc{xabs3} & $N_{\mathrm{H}} / 10^{20}~\mathrm{cm^{-2}}$ & \multicolumn{5}{c}{$6.6^{+1.6}_{-2.5}$} \\
			& $\log\xi$ & \multicolumn{5}{c}{$<-1.8$} \\
			& $v_{\mathrm{turb}}/\mathrm{km~s^{-1}}$ & \multicolumn{5}{c}{$110^f$} \\			
			& $v_{\mathrm{outflow}}/\mathrm{km~s^{-1}}$ & \multicolumn{5}{c}{$140^f$} \\
			& $z$ & \multicolumn{5}{c}{$0.0604^f$} \\

			$\times($ & & & & & & \\			

			\textsc{cflux1} & $E_{\mathrm{min}}/\mathrm{keV}$ & \multicolumn{5}{c}{$0.1^{f}$} \\
			& $E_{\mathrm{max}}/\mathrm{keV}$ & \multicolumn{5}{c}{$100^{f}$} \\
			& $\log F$ & $-10.54\pm0.02$ & $-10.41^{+0.01}_{-0.02}$ & $-10.58\pm0.02$ & $-10.51\pm0.02$ & $-10.51\pm0.02$ \\

			$\times$ & & & & & & \\

			\textsc{cutoffpl} & $\Gamma$ & $2.32\pm0.02$ & $2.29\pm0.02$ & $2.30\pm0.02$ & $2.39\pm0.02$ & $2.31\pm0.02$ \\
			& $E_{\mathrm{cut}}/\mathrm{keV}$ & \multicolumn{5}{c}{$300^{f}$} \\

			$+$ & & & & & & \\

			\textsc{const} & $R$ & $0.49^{+0.15}_{-0.09}$ & $0.45^{+0.14}_{-0.08}$ & $0.73\pm0.20$ & $0.72^{+0.20}_{-0.15}$ & $0.68^{+0.20}_{-0.13}$ \\

			$\times$ & & & & & & \\

			\textsc{cflux2} & $E_{\mathrm{min}}/\mathrm{keV}$ & \multicolumn{5}{c}{$E^{l}_{\mathrm{min},~\textsc{cflux1}}$} \\
			& $E_{\mathrm{max}}/\mathrm{keV}$ & \multicolumn{5}{c}{$E^{l}_{\mathrm{max},~\textsc{cflux1}}$} \\
			& $\log F$ & \multicolumn{5}{c}{$\log F^{l}_{\textsc{cflux1}}$} \\

			$\times$ & & & & & & \\

			\textsc{relxilld} & $q_{\mathrm{in}}$ & \multicolumn{5}{c}{$7.0^{+0.6}_{-0.2}$} \\
			& $q_{\mathrm{out}}$ & \multicolumn{5}{c}{$3^{f}$} \\
			& $R_{\mathrm{break}}/r_g$ & \multicolumn{5}{c}{$6^{f}$} \\
			& $a$ & \multicolumn{5}{c}{$0.998^{f}$} \\
			& $i/^{\circ}$ & \multicolumn{5}{c}{$24^{+3}_{-4}$} \\
			& $R_{\mathrm{in}}/r_g$ & \multicolumn{5}{c}{$2.5^{+0.4}_{-0.2}$} \\
			& $R_{\mathrm{out}}/r_g$ & \multicolumn{5}{c}{$400^{f}$} \\
			& $z$ & \multicolumn{5}{c}{$0.0604^{f}$} \\
			& $\Gamma$ & \multicolumn{5}{c}{$\Gamma^{l}_{\textsc{cutoffpl}}$} \\
			& $\log \xi$ & \multicolumn{5}{c}{$2.71^{+0.06}_{-0.01}$} \\
			& $A_{\mathrm{Fe}}$ & \multicolumn{5}{c}{$1^f$} \\
			& $\log N/~\mathrm{cm^{-3}}$ & \multicolumn{5}{c}{$>18.9$} \\

			$+$ & & & & & & \\

			\textsc{zashift} & $z$ & \multicolumn{5}{c}{$0.0604^{f}$} \\

			$\times($ & & & & & & \\

			\textsc{laor} & $E_{\mathrm{rest}}/\mathrm{keV}$ & \multicolumn{5}{c}{$7.1\pm0.1$} \\
			& $q$ & \multicolumn{5}{c}{$3^{f}$} \\
			& $R_{\mathrm{in}}/r_g$ & \multicolumn{5}{c}{$10^{+10}_{-4}$} \\
			& $R_{\mathrm{out}}/r_g$ & \multicolumn{5}{c}{$R^{l}_{\mathrm{out},~\textsc{relxilld}}$} \\
			& $i/^{\circ}$ & \multicolumn{5}{c}{$i^{l}_{\textsc{relxilld}}$} \\
			& $N/10^{-5}~\mathrm{photons~s^{-1}~cm^{-2}}$ & \multicolumn{5}{c}{$1.3\pm0.3$} \\

			$))$ & & & & & & \\
			\hline
			\label{tab:params-final}
		\end{tabular}
	\end{center}
\end{table*}

\begin{figure*}
	\scalebox{1.0}{\includegraphics[width=\linewidth]{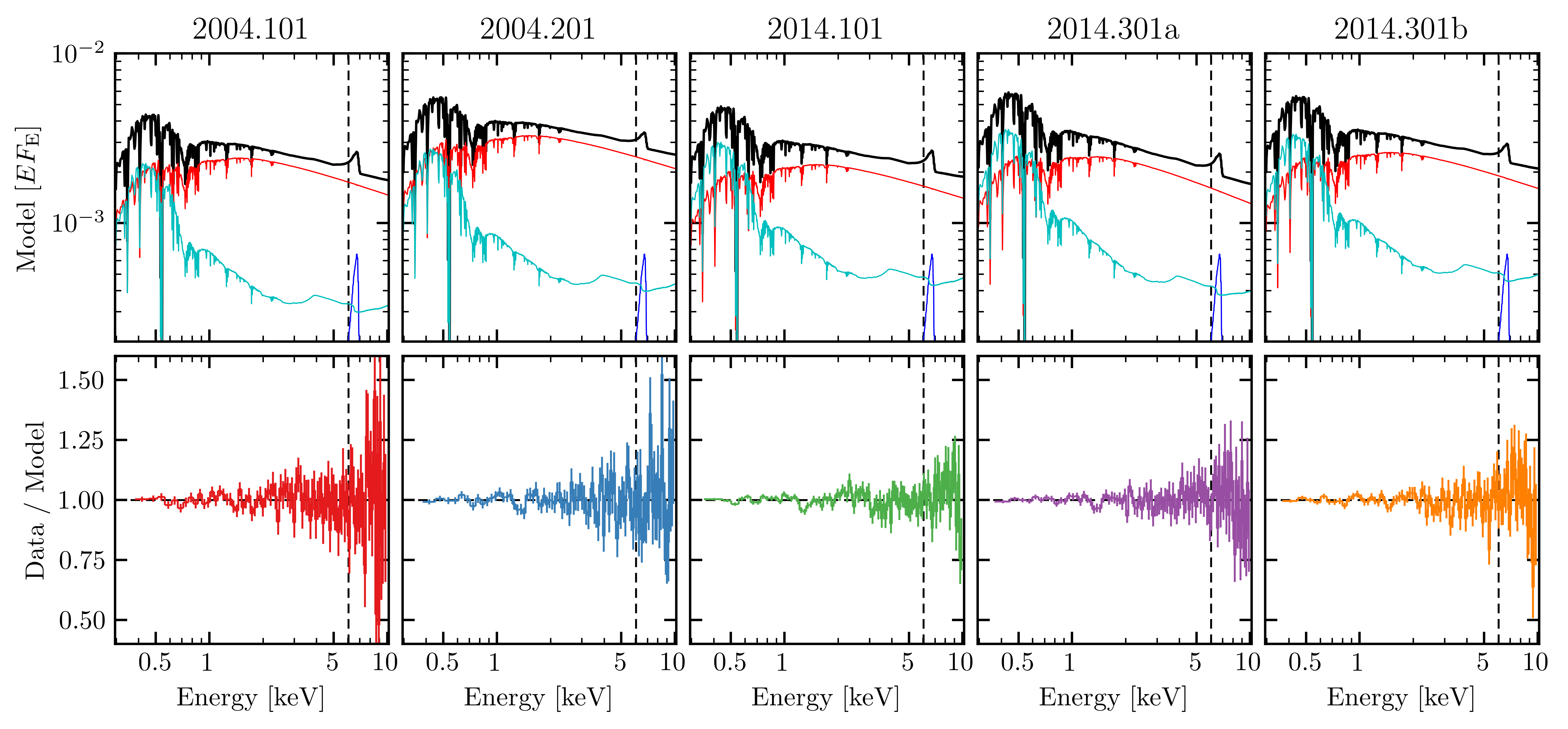}}
	\caption{\textit{Top row} $-$ Total model (black) and model components: power-law (red), reflection (cyan), and broad Fe emission line (blue). The curves are plotted in $EF_{\mathrm{E}}$ units of $\mathrm{keV^{2}~photons~cm^{-2}~s^{-1}~keV^{-1}}$. \textit{Bottom row} $-$ Coloured (grey) curves denote the ratio residuals of the total (background) spectrum in each epoch when divided by the model displayed in the top row of the corresponding column. Vertical lines in each panel denote the observed frame Fe K$\alpha$ emission line energy, which has a rest frame energy of $6.4~\mathrm{keV}$.}
	\label{fig:plra-final}
\end{figure*}

We therefore arrived at two best-fit models, one in which low-energy deviations are accounted for by allowing the warm absorber parameters to be free to vary and one in which excess Fe L emission is present. In order to evaluate the goodness-of-fit for each scenario we used the method presented in \cite{Kaastra2017}, computing the expected $C$-statistic and its variance $\left(C_{\mathrm{E}},~C_{\mathrm{V}}\right)$, where we found that $\left(C_{\mathrm{E}},~C_{\mathrm{V}}\right) = \left(907\pm13.5\right)$ for the 10 simultaneously fit spectra. The excess Fe L emission scenario resulted in $C = 871$ for $876$ degrees of freedom, while the free to vary warm absorber model returned $C = 900$ for $878$ degrees of freedom. Comparing these to the expected $C$-statistic and its variance finds that the inclusion of Fe L emission can be rejected at a $2\sigma$ level, whereas the varying warm absorber model cannot be rejected even at the $1\sigma$ level. We report this final best-fit model and its parameters in Table \ref{tab:params-final}, with the ratio residuals shown in Figure \ref{fig:plra-final}. 

Despite achieving a statistically acceptable fit to the data, significant high-energy residuals remain throughout the epochs. Especially notable are the residuals in the 2014 epochs, which all show slight excesses at $\sim8~\mathrm{keV}$ before decreasing toward higher energies. Strong emission features that would account for this curvature are not known, and were therefore not explored. Blue-shifted Fe K absorption features at energies $>7.1~\mathrm{keV}$ have been reported in numerous AGN (e.g. \citealt{Tombesi+2010,Gofford+2013,Igo+2020}), and indeed evidence of ultra-fast outflows from other RGS based analyses (\citealt{Longinotti+2015,Sanfrutos+2018}) having already been detected in this source with velocities of $\sim0.05-0.1c$. \cite{Longinotti+2015} reported finding marginal evidence of blue-shifted Fe K absorption that would correspond to an outflow of $\sim0.3c$. We performed a search for such features in each epoch by stepping an inverted narrow (i.e. $\sigma=10~\mathrm{eV}$) Gaussian component between $7.1-10.5~\mathrm{keV}$ in the source frame, allowing energy and normalisation to be free to vary. We obtained similar absorption feature energies as \cite{Longinotti+2015}, though we found a statistically insignificant fit improvement for their inclusion ($\Delta C = -14$ for $10$ additional free parameters) and therefore do not include them in the reported model.

The covariance spectra produced in Section \ref{sect:timing} were then explored quantitatively by using the reported best-fit broad band spectral model. Covariance spectra for the low- and high-frequency combined analysis were loaded into \textsc{xspec} via the \textsc{ftflx2xsp} tool after accounting for detector effective area in each energy band, and subsequently fit with the model in Table \ref{tab:params-final}. During the fitting procedure we initially allowed only the normalisation of each component to be free to vary, which showed that the power-law component was sufficient on its own in fitting the the covariance spectra. This fit, however, was quite poor ($\chi^2/dof \approx 1.5$) in describing both data sets and therefore we allowed the power-law photon index to be free to vary between the two frequency regimes, finding the data to be over fit ($\chi^2/dof \approx 0.7$) but with residuals significantly improved across all energies. The low-frequency data were best-fit by a power-law with $\Gamma = 2.37\pm0.12$ while the high-frequency data returned $\Gamma = 2.02\pm0.07$. This suggests a possible two-corona scenario where the harder power-law varies on shorter timescales, corresponding to a compact coronal structure, than a more slowly varying soft power-law, corresponding to a more extended geometry. The absence of a detected reflection component suggests that it may be varying on significantly longer timescales than those explored here. Fitting the time-resolved high-frequency covariance spectra returned the same hard power-law as the combined analysis, albeit with varying normalisations.
	
With this interesting result we attempted to fit our broad band data sets with a two-corona scenario by restricting the power-law photon indices to be within $\pm10~\mathrm{per~cent}$ of the aforementioned values. Doing so found no statistically significant fit improvement when including the hard power-law from the high-frequency covariance fit, recovering only the soft power-law from the low-frequency covariance fit. Indeed this soft power-law agrees very well with the best-fit model presented in Table \ref{tab:params-final}, suggesting it dominates the observed emission.

We note in Section \ref{sect:timing} that the changing lag-frequency spectrum observed during the 2014.301a epoch may be interpreted as evidence of a moving corona, specifically an increase in height above the accretion disc. Ray-tracing simulations exploring accretion discs illuminated by compact coronal geometries predict a decrease in $q_{\mathrm{in}}$ for an increase in source height (e.g. \citealt{WilkinsFabian2012,Gonzalez+2017}), however, we found no statistically significant fit improvement when allowing $q_{\mathrm{in}}$ for the 2014.301a epoch to be free to vary independently of the other epochs.

\section{Discussion}
\label{sect:discussion}
In this work we have presented the first detailed timing and spectral analyses on the four \xmm EPIC pn observations of the RL-NLS1 \src. We have found that the source flux varies by $\sim35~\mathrm{per~cent}$ and the hardness ratio by $\sim20~\mathrm{per~cent}$ over the $10~\mathrm{yr}$ period spanned by the observations. We note that the flux levels reported here in Table \ref{tab:lcfits} are essentially unchanged from those measured by \cite{LeighlyTiming1999} using \asca data from 1995, revealing low levels of long-term variability over $\sim20~\mathrm{yr}$. These characteristics are not extreme when compared to other NLS1s, such as Mrk 335 \citep{Gallo+2018} and IRAS 13224$-$3809 \citep{Alston+2019}, which have been shown to exhibit significant short- and long-term variability. The first $40~\mathrm{ks}$ of the second 2014 observation were found to be significantly softer than at any other time, prompting the segmentation of this observation into two epochs in which separate timing and spectral analyses were performed in an effort to uncover the cause of such a spectral state change. 

Further abnormalities were found in the behaviour of \src with respect to its variability properties when compared to the bulk of its RQ-NLS1 counterparts, displaying a `harder-when-brighter' trend rather than `softer-when-brighter', increasing fractional variability with energy rather than being peaked at low energies or flat, and a possible non-stationary process evident only at low energies. The `harder-when-brighter' trend observed here is similar to the trends observed in I Zw 1 (RQ-NLS1; \citealt{Gallo+2004,Gallo+2007}) and IRAS 16318$-$472 (RL-NLS1; \citealt{Mallick+2016}), where the authors also reported similar trends of increasing fractional variability with energy in both sources, leading to a suggested variable power-law in both photon index and normalisation to explain the variability in those objects. Our flux-flux analysis of the various observation epochs of \src agrees with this simple variable power-law explanation. We note, however, that the observed `harder-when-brighter' trend is generally more often observed in blazars (e.g. \citealt{Bhatta+2018,Singh+2019,Zhang+2019}), particularly during flaring episodes, than in Seyfert-type AGN. Since \src and IRAS 16318$-$472 are both jetted sources it is plausible that the jet emission is influencing and/or connected to the X-ray emission. Indeed an aborted jet scenario was proposed for I Zw 1 \citep{Gallo+2007,Wilkins+2017} wherein flaring of the hard X-ray corona was best explained as an episode of vertical collimation and outflow, which may be interpreted as the X-ray corona acting as the base of a jet. Moreover, \cite{Foschini+2015} found that when normalising by black hole mass the jet properties of RL-NLS1s aligned well with those of flat-spectrum radio quasars (FSRQs) and BL Lac objects, appearing to be the low-power tail of the more powerful, higher black hole mass blazar-type sources. 

Frequency-domain methods applied to the light curves revealed many similarities with other RQ-NLS1s, with the PSD here being well fit by a simple power-law description, though no break frequency could be constrained, and the lag-frequency spectra exhibiting a strong high-frequency reverberation signature, providing evidence of X-ray reflection that is frequently found in other NLS1s (e.g. \citealt{Zoghbi+2010,Kara+2016}). Using this high-frequency reverberation lag we estimated the mass of the SMBH in \src using the relationships derived in \cite{Demarco+2013} relating the frequency of the reverberation lag and the absolute value of the reverberation lag to the black hole mass. The lag-frequency spectrum in Figure \ref{fig:cohlag} has a clearly defined negative lag at $\left(\nu,~\left|\tau\right|\right) = \left(2.12\times10^{-4}~\mathrm{Hz},~270~\mathrm{s}\right)$, allowing for the black hole mass estimation to be performed using the aforementioned $\nu-M_{\mathrm{BH}}$ and $\left|\tau\right|-M_{\mathrm{BH}}$ relations, finding $2.34\times10^7~M_{\odot}$ and $5.82\times10^7~M_{\odot}$, respectively. Previous $M_{\mathrm{BH}}$ estimates for \src have been as low as $3.1\times10^6~M_{\odot}$ \citep{Berton+2015} to as high as $5.4\times10^7 M_{\odot}$ \citep{Berton+2016}, with \cite{Nikolajuk+2009} finding $1.08\times10^7~M_{\odot}$. Our estimates here therefore broadly agree with the high-mass estimates of previous studies, which themselves are typical mass estimates for NLS1s, usually on the order of a few $\times10^7~M_{\odot}$ (e.g. \citealt{Berton+2015}). We then used the empirical relationship derived by \cite{McHardy+2006} between bolometric luminosity and black hole mass to estimate the PSD break frequency. The mean of our mass estimates calculated here via the combined lag-frequency spectrum is $M_{\mathrm{BH}} = 4.08\times10^7~M_{\odot}$. Using the \textsc{lumin} function in \textsc{xspec} on our best-fit spectral model we found $L_{\mathrm{2-10}} = 5.84\times10^{43}~\mathrm{erg~s^{-1}}$. \cite{Duras+2020arXiv} recently reported bolometric corrections to the X-ray luminosity including those based on $M_{\mathrm{BH}}$, which can be used to estimate this value as $L_{\mathrm{bol}} = 16.86 \times L_{\mathrm{2-10}} = 9.85\times10^{44}~\mathrm{erg~s^{-1}}$. These quantities provide an estimated PSD break frequency of $\nu_{\mathrm{break}} \approx 9.43\times10^{-6}~\mathrm{Hz}$, which is below the lowest sampled frequency available with these data, explaining why we were unable to properly constrain this parameter during the attempted bending power-law fits. 

The timing analysis results provided strong evidence of a variable power-law spectral component in the presence of some reflection component, motivating the use of such a two-component model in the spectral modelling. We found the five epochs to be best-fit by a scenario in which the power-law component was variable in both normalisation and photon index across all epochs, with its emission being reflected off of a highly ionised, high density accretion disc truncated just outside the ISCO of a maximally spinning Kerr black hole. Excess residuals were still present in the Fe K and Fe L energy bands, with the broad Fe K emission feature being well fit by a Gaussian feature with similar parameters to those measured by \cite{LeighlySpectral1999} using \textit{ASCA} data from 1995, revealing a persistent, broad Fe K emission line. When fitting this feature with a Gaussian line profile we found $E\approx6.8~\mathrm{keV}$ and $\sigma\approx0.5~\mathrm{keV}$, translating into a FWHM of $v\approx50,000~\mathrm{km~s^{-1}}$, which would correspond to $r=GM/v^2\approx35~r_g$ assuming that the emission is produced by an accretion disc `hot-spot' on a Keplerian orbit around the central SMBH.

We attempted to fit the residuals using the \textsc{diskline} model which revealed an emission region consistent with an annulus between $10-45~r_g$, though these values were not well constrained. Using our lag-derived mean $M_{\mathrm{BH}}$ estimate we approximate an expected reverberation lag of $\sim7~\mathrm{ks}$ for material at this radius, however, the high-frequency lag-energy spectrum of Figure \ref{fig:lag-energy} reveals no such lag at the corresponding energy.

The absence of a detected reverberation lag in the $5-7~\mathrm{keV}$ band is puzzling as it suggests that the emission feature is not responding to continuum variations. Physically this may correspond to a simple scenario where the emission region is in reality significantly further away from the inner disc region than our derived emission line parameters would suggest. Indeed it was found that a distant reflection scenario (i.e. off of the torus) fits a significant portion of the Fe K band deviations from our intermediate continuum fit, however, a broad emission line provided a substantial statistical improvement over this scenario. Alternatively, the ultra-fast outflows detected in the RGS spectra of this source may produce a shielding wind structure that prevents the coherent propagation of continuum variations to the inferred `hot-spot' radius. It is also possible that the origin of the excess Fe K band emission is due a more complex reflection scenario in which the reflection spectrum of the accretion disc is Comptonised by an extended corona covering the inner disc (e.g. \citealt{Petrucci+2001,WilkinsGalloCOMP2015}). \cite{WilkinsGalloCOMP2015} showed that Comptonisation of the accretion disc reflection spectrum by an extended patchy corona can produce an enhanced blue `horn' of the Fe K$\alpha$ emission feature while also artificially increasing the recovered inner disc radius relative to the input value on the simulated spectrum. Furthermore, the ionisation state the disc material is expected to follow a gradient which may be poorly fit when assuming a fixed value for the entirety of the disc, as was done here in using the \textsc{relixlld} model. The origin of the Fe K excess emission remains difficult to constrain based on the currently available data.

As described in Section \ref{sect:spectral} the best-fit model was achieved by allowing the RGS-derived warm absorber parameters to be free to vary, with the excess Fe L emission scenario ruled out when evaluating the goodness-of-fit. Despite this, we tested the validity of such a scenario by comparing our observed line normalisation ratio $N_{\mathrm{L}} / N_{\mathrm{K}} = \left(L/K\right)_{\mathrm{O}}$ to the theoretical predictions of \cite{Kallman1995}, where it was found that for a gas with ionisation parameter $\log \xi \approx 2.5$ being photoionised by a power-law with photon index $\Gamma = 2.5$ the theoretically expected line ratio is $\left(L/K\right)_{\mathrm{T}}\approx1.6$. When we inserted an additional Fe L feature without allowing the warm absorbers to be free to vary we found that the observed line ratio was $\left(L/K\right)_{\mathrm{O}} = 3.2 \pm 0.6$, and when the warm absorber parameters were freed this changed to $\left(L/K\right)_{\mathrm{O}} = 2.9 \pm 0.8$. In both cases the line ratio does not agree with the theoretically predicted value. We also note that the best-fit line energies of these Fe K and Fe L features are significantly higher than their rest-frame energies and not as theoretically predicted by \cite{Kallman1995}. This result may by due to the asymmetry of the blurred line profile which was added after the power-law and blurred ionised reflector had fit the continuum. These components may have already fit the present red-shifted `wing' of the line, leaving only the blue-shifted `horn' of the feature to be fit by the separate emission features described here, resulting in the higher line energies reported.

The nature of the variability was further explored by applying our best-fit spectral model to the covariance spectra, where we found that the low-frequency spectrum was well described by a soft power-law ($\Gamma\approx2.4$) while the high-frequency spectrum required a significantly harder power-law ($\Gamma\approx2.0$). This is interpreted as evidence of a two-corona scenario, in which a compact, hard corona varies more rapidly than an extended, soft corona that varies more slowly. Evidence of this complex coronal geometry has been reported in RQ-NLS1s such as 1H~0707$-$495 \citep{Kara+2013} as well as in the flaring states of I Zw 1 \citep{Gallo+2007,Wilkins+2017} and Mrk 335 \citep{WilkinsGallo2015}. This extended coronal geometry may in fact be responsible for producing the observed excess Fe K band emission via the aforementioned Comptonisation of the accretion disc reflection spectrum. Here we are unable to significantly detect the hard power-law component in the spectrum, perhaps indicating that it is significantly weaker than the soft power-law. It may also be true that the currently available data are not of sufficient quality to apply such a model consisting of a double power-law plus highly ionised, relativistically blurred reflection spectrum modified by multiple warm absorbers, where there are significant degeneracies between the numerous free parameters of the multiple spectral components.

\begin{figure}
	\scalebox{1.0}{\includegraphics[width=\linewidth]{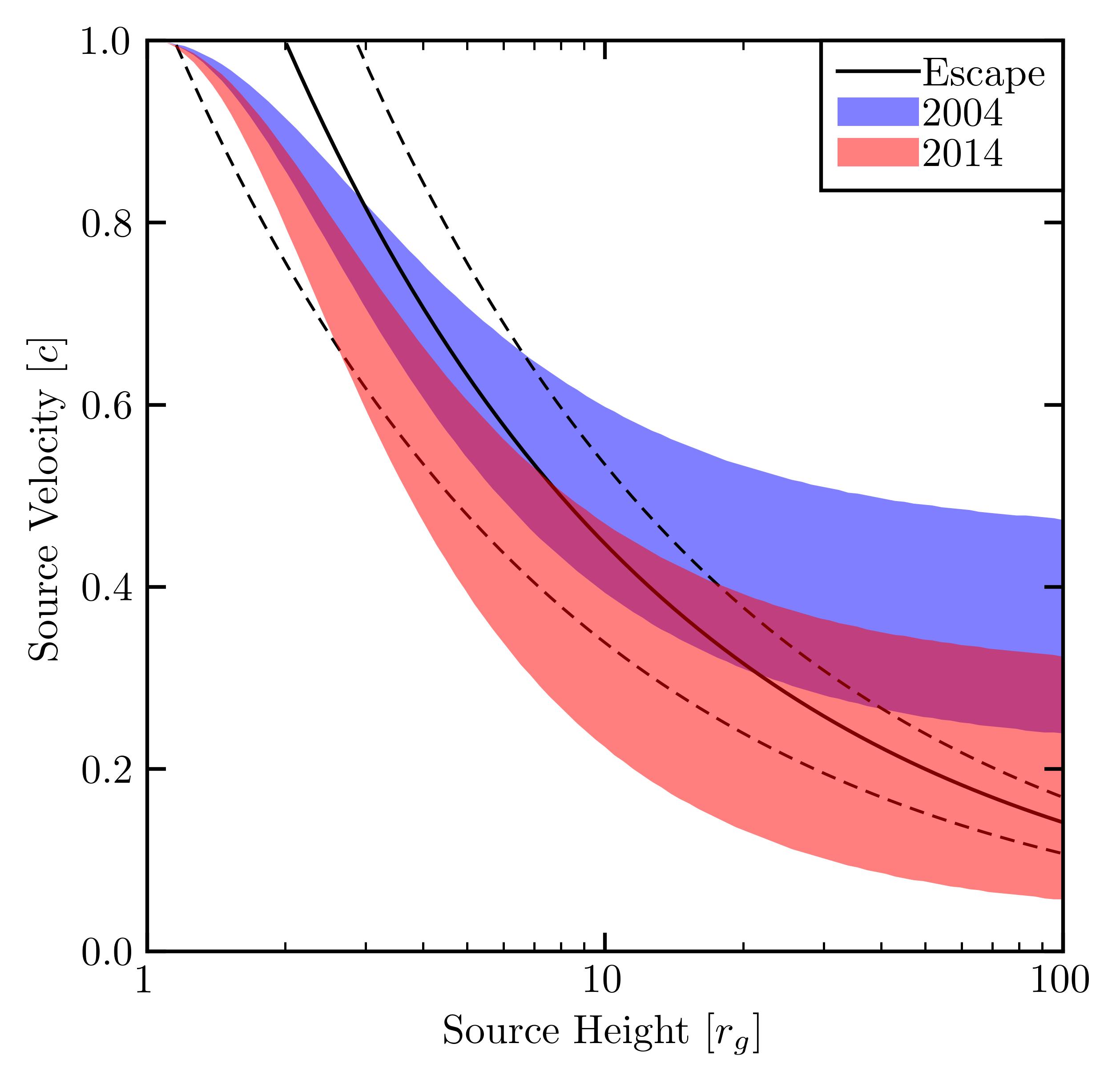}}
	\caption{The coronal parameters required to enable escape from the black hole environment. The black solid line represents the velocity needed at each height to escape the gravitational potential of a maximal spin ($a=0.998$) Kerr black hole of mass $\left(4.08\pm1.74\right)\times10^7~M_{\odot}$, with the dashed lines corresponding to the lower and upper mass limits. The blue and red filled regions represent the source parameters estimated using the range of reflection fraction values obtained from the spectral fits for the 2004 ($R = 0.47^{+0.17}_{-0.10}$) and 2014 ($R = 0.71^{+0.22}_{-0.18}$) data sets, respectively.}
	\label{fig:coronaparams}
\end{figure}

The radio emission in this source indicates the presence of a radio jet, although whether it is a sub-relativistic or misaligned jet remains unknown \citep{Doi+2011,Giroletti+2017}. \cite{Gonzalez+2017} showed that for beamed X-ray coronae the reflection fraction will be $R<1$ and presented a method to compute the height and velocity of the outflowing corona given the reflection fraction from spectral modelling. The best-fit model parameters in Table \ref{tab:params-final} therefore suggest a beamed corona in \src. Considering that the 2004 and 2014 best-fit parameters give different reflection fraction values between the two observing periods, but are consistent with epochs within those periods, we took the mean reflection fraction values for each, giving $R = 0.47^{+0.17}_{-0.10}$ for 2004 and $R = 0.71^{+0.22}_{-0.18}$ for 2014. We compared these source parameter curves to the escape velocity required to overcome the gravitational potential of the black hole in \src by taking the mean of the mass estimates obtained from our combined lag-frequency spectrum as $M_{\mathrm{BH}} = \left(4.08\pm1.74\right)\times10^7~M_{\odot}$. This revealed the combinations of source parameters that will result in an observable outflow. The results are shown in Figure \ref{fig:coronaparams}, where the X-ray corona is shown to be capable of outflowing from the system for source heights as close as $\sim3~r_g$ in 2004 but at larger heights of $\sim7~r_g$ in 2014. If we invoke the X-ray corona as the base of the radio jet then these results can be interpreted in two ways: (i) the X-ray source has increased in height between the two epochs, or (ii) the X-ray source has significantly slowed over the $10~\mathrm{yr}$ period and maintained a constant source height. These two scenarios would have very different impacts on the emissivity profile of the accretion disc. For example, in case (i) the source has increased in height, resulting in less emission incident on the inner disc, thereby lowering $q_{\mathrm{in}}$. In case (ii), the source will have become more weakly collimated, thereby allowing more emission to be gravitationally bent back toward the accretion disc thus increasing $q_{\mathrm{in}}$. When attempting to fit such scenarios to the data here we were unsuccessful in sufficiently constraining the parameters such that meaningful statistical differences could be determined, and thus they are not presented here. We note that the sub-relativistic scenario (i.e. $\beta \lesssim 0.9c$) suggested in previous radio studies agrees with the results here, where the maximum velocity attained resulting in an outflow is $\sim0.8c$.

\section{Conclusion}
\label{sect:conclusion}
We found that IRAS 17020+4544, a RL-NLS1 with a sub-relativistic radio jet, is in many ways comparable to RQ counterparts, while also showing significant differences regarding variability trends. 

Notable differences include a `harder-when-brighter' trend, increasing variability with energy, and evidence of possible non-stationarity at low energies. The hardness-flux relationship and fractional variability spectra are found to be similar to other sources, both RQ and RL, that show evidence of jet activity. In such sources a variable power-law, both in normalisation and photon index, is shown to be sufficient in explaining the observed trends.

Similarities with RQ sources are found with the confirmation of a reverberation lag, providing evidence of significant disc reflection in the X-ray spectrum of a RL source, as well as a standard PSD that can be well described by a simple power-law. Computation of the low- and high-frequency covariance spectra suggests that a variable power-law dominates the observed flux variability across a wide range of time-scales, with a compact hard corona being variable on shorter timescales than an extended soft corona which varies on longer timescales. 

The evolution of all timing products as the source enters a soft spectral state was explored, finding significantly increased variability across all energies, in both the fractional variability and covariance spectra, as well as a shifted lag-frequency spectrum to lower frequencies. These pieces of evidence indicate a significant spectral state change, potentially explained by a moving X-ray corona.

Simultaneous spectral modelling of all five epochs was performed with parameters motivated by the timing results. We find that a variable power-law in the presence of a relativistically blurred, highly ionised reflector is sufficient in explaining the X-ray spectra. The spectra also display evidence of significant excess Fe K emission beyond the included reflection model, perhaps from an orbiting hot-spot in the accretion disc at $r\approx35~r_g$.

\src provides another look into a very limited class of AGN: a RL, jetted NLS1 with strong disc emission in the X-ray spectrum. Such sources provide important testing grounds for disc-jet connection theories. Here we show that, as in a handful of other NLS1s, the power-law spectral component produced by the X-ray corona may be linked to the radio jet base.

\section*{Acknowledgements}
We thank the anonymous referee for their helpful comments which improved the overall quality of the manuscript. AGG and LCG acknowledge financial support from the Natural Sciences and Engineering Research Council of Canada (NSERC). LCG acknowledges financial support from the Canadian Space Agency (CSA).

\section*{Data Availability}
The archival data used in this work are publicly available and can be downloaded from the \xmm Science Archive (XSA) and were processed using the \xmm Science Analysis System (SAS). We made use of the software packages \textsc{heasoft} and \textsc{xspec} which can be downloaded from the High Energy Astrophysics Science Archive Research Centre (HEASARC) software web-page. Further analysis and figure production were performed using \textsc{r} version 3.5 and \textsc{python} version 3.6.



\bibliographystyle{mnras}
\bibliography{refs} 

\begin{thebibliography}{}
\makeatletter
\relax
\def\mn@urlcharsother{\let\do\@makeother \do\$\do\&\do\#\do\^\do\_\do\%\do\~}
\def\mn@doi{\begingroup\mn@urlcharsother \@ifnextchar [ {\mn@doi@}
  {\mn@doi@[]}}
\def\mn@doi@[#1]#2{\def\@tempa{#1}\ifx\@tempa\@empty \href
  {http://dx.doi.org/#2} {doi:#2}\else \href {http://dx.doi.org/#2} {#1}\fi
  \endgroup}
\def\mn@eprint#1#2{\mn@eprint@#1:#2::\@nil}
\def\mn@eprint@arXiv#1{\href {http://arxiv.org/abs/#1} {{\tt arXiv:#1}}}
\def\mn@eprint@dblp#1{\href {http://dblp.uni-trier.de/rec/bibtex/#1.xml}
  {dblp:#1}}
\def\mn@eprint@#1:#2:#3:#4\@nil{\def\@tempa {#1}\def\@tempb {#2}\def\@tempc
  {#3}\ifx \@tempc \@empty \let \@tempc \@tempb \let \@tempb \@tempa \fi \ifx
  \@tempb \@empty \def\@tempb {arXiv}\fi \@ifundefined
  {mn@eprint@\@tempb}{\@tempb:\@tempc}{\expandafter \expandafter \csname
  mn@eprint@\@tempb\endcsname \expandafter{\@tempc}}}

\bibitem[\protect\citeauthoryear{{Alston}}{{Alston}}{2019}]{Alston2019}
{Alston} W.~N.,  2019, \mn@doi [\mnras] {10.1093/mnras/stz423}, \href
  {https://ui.adsabs.harvard.edu/abs/2019MNRAS.485..260A} {485, 260}

\bibitem[\protect\citeauthoryear{{Alston} et~al.,}{{Alston}
  et~al.}{2019}]{Alston+2019}
{Alston} W.~N.,  et~al., 2019, \mn@doi [\mnras] {10.1093/mnras/sty2527}, \href
  {https://ui.adsabs.harvard.edu/abs/2019MNRAS.482.2088A} {482, 2088}

\bibitem[\protect\citeauthoryear{{Alston} et~al.,}{{Alston}
  et~al.}{2020}]{Alston+2020}
{Alston} W.~N.,  et~al., 2020, \mn@doi [Nature Astronomy]
  {10.1038/s41550-019-1002-x}, \href
  {https://ui.adsabs.harvard.edu/abs/2020NatAs.tmp....2A} {p.~2}

\bibitem[\protect\citeauthoryear{{Anderson} \& {Darling}}{{Anderson} \&
  {Darling}}{1952}]{AndersonDarling1952}
{Anderson} T.~W.,  {Darling} D.~A.,  1952, \mn@doi [Ann. Math. Statist.]
  {10.1214/aoms/1177729437}, 23, 193

\bibitem[\protect\citeauthoryear{{Ar{\'e}valo} \& {Uttley}}{{Ar{\'e}valo} \&
  {Uttley}}{2006}]{ArevaloUttley2006}
{Ar{\'e}valo} P.,  {Uttley} P.,  2006, \mn@doi [\mnras]
  {10.1111/j.1365-2966.2006.09989.x}, \href
  {https://ui.adsabs.harvard.edu/abs/2006MNRAS.367..801A} {367, 801}

\bibitem[\protect\citeauthoryear{{Arnaud}}{{Arnaud}}{1996}]{Arnaud1996}
{Arnaud} K.~A.,  1996, in {Jacoby} G.~H.,  {Barnes} J.,  eds,  Astronomical
  Society of the Pacific Conference Series Vol. 101, Astronomical Data Analysis
  Software and Systems V. p.~17

\bibitem[\protect\citeauthoryear{{Berton} et~al.,}{{Berton}
  et~al.}{2015}]{Berton+2015}
{Berton} M.,  et~al., 2015, \mn@doi [\aap] {10.1051/0004-6361/201525691}, \href
  {https://ui.adsabs.harvard.edu/abs/2015A&A...578A..28B} {578, A28}

\bibitem[\protect\citeauthoryear{{Berton}, {Foschini}, {Ciroi}, {Cracco}, {La
  Mura}, {Di Mille}  \& {Rafanelli}}{{Berton} et~al.}{2016}]{Berton+2016}
{Berton} M.,  {Foschini} L.,  {Ciroi} S.,  {Cracco} V.,  {La Mura} G.,  {Di
  Mille} F.,   {Rafanelli} P.,  2016, \mn@doi [\aap]
  {10.1051/0004-6361/201527056}, \href
  {https://ui.adsabs.harvard.edu/abs/2016A&A...591A..88B} {591, A88}

\bibitem[\protect\citeauthoryear{{Berton} et~al.,}{{Berton}
  et~al.}{2018}]{Berton+2018}
{Berton} M.,  et~al., 2018, \mn@doi [\aap] {10.1051/0004-6361/201832612}, \href
  {https://ui.adsabs.harvard.edu/abs/2018A&A...614A..87B} {614, A87}

\bibitem[\protect\citeauthoryear{{Bhatta}, {Mohorian}  \& {Bilinsky}}{{Bhatta}
  et~al.}{2018}]{Bhatta+2018}
{Bhatta} G.,  {Mohorian} M.,   {Bilinsky} I.,  2018, \mn@doi [\aap]
  {10.1051/0004-6361/201833628}, \href
  {https://ui.adsabs.harvard.edu/abs/2018A&A...619A..93B} {619, A93}

\bibitem[\protect\citeauthoryear{{Boller}, {Brandt}  \& {Fink}}{{Boller}
  et~al.}{1996}]{Boeller+1996}
{Boller} T.,  {Brandt} W.~N.,   {Fink} H.,  1996, \aap, \href
  {https://ui.adsabs.harvard.edu/abs/1996A&A...305...53B} {305, 53}

\bibitem[\protect\citeauthoryear{{Boroson} \& {Green}}{{Boroson} \&
  {Green}}{1992}]{BorosonGreen1992}
{Boroson} T.~A.,  {Green} R.~F.,  1992, \mn@doi [\apjs] {10.1086/191661}, \href
  {https://ui.adsabs.harvard.edu/abs/1992ApJS...80..109B} {80, 109}

\bibitem[\protect\citeauthoryear{{Brandt}, {Mathur}  \& {Elvis}}{{Brandt}
  et~al.}{1997}]{Brandt+1997}
{Brandt} W.~N.,  {Mathur} S.,   {Elvis} M.,  1997, \mn@doi [\mnras]
  {10.1093/mnras/285.3.L25}, \href
  {https://ui.adsabs.harvard.edu/abs/1997MNRAS.285L..25B} {285, L25}

\bibitem[\protect\citeauthoryear{{Cash}}{{Cash}}{1979}]{Cash1979}
{Cash} W.,  1979, \mn@doi [The Astrophysical Journal] {10.1086/156922}, \href
  {https://ui.adsabs.harvard.edu/abs/1979ApJ...228..939C} {228, 939}

\bibitem[\protect\citeauthoryear{{De Marco}, {Ponti}, {Cappi}, {Dadina},
  {Uttley}, {Cackett}, {Fabian}  \& {Miniutti}}{{De Marco}
  et~al.}{2013}]{Demarco+2013}
{De Marco} B.,  {Ponti} G.,  {Cappi} M.,  {Dadina} M.,  {Uttley} P.,  {Cackett}
  E.~M.,  {Fabian} A.~C.,   {Miniutti} G.,  2013, \mn@doi [\mnras]
  {10.1093/mnras/stt339}, \href
  {https://ui.adsabs.harvard.edu/abs/2013MNRAS.431.2441D} {431, 2441}

\bibitem[\protect\citeauthoryear{{Doi}, {Asada}  \& {Nagai}}{{Doi}
  et~al.}{2011}]{Doi+2011}
{Doi} A.,  {Asada} K.,   {Nagai} H.,  2011, \mn@doi [\apj]
  {10.1088/0004-637X/738/2/126}, \href
  {https://ui.adsabs.harvard.edu/abs/2011ApJ...738..126D} {738, 126}

\bibitem[\protect\citeauthoryear{{Done}, {Davis}, {Jin}, {Blaes}  \&
  {Ward}}{{Done} et~al.}{2012}]{Done+2012}
{Done} C.,  {Davis} S.~W.,  {Jin} C.,  {Blaes} O.,   {Ward} M.,  2012, \mn@doi
  [\mnras] {10.1111/j.1365-2966.2011.19779.x}, \href
  {https://ui.adsabs.harvard.edu/abs/2012MNRAS.420.1848D} {420, 1848}

\bibitem[\protect\citeauthoryear{{Duras} et~al.,}{{Duras}
  et~al.}{2020}]{Duras+2020arXiv}
{Duras} F.,  et~al., 2020, arXiv e-prints, \href
  {https://ui.adsabs.harvard.edu/abs/2020arXiv200109984D} {p. arXiv:2001.09984}

\bibitem[\protect\citeauthoryear{{Foschini} et~al.,}{{Foschini}
  et~al.}{2015}]{Foschini+2015}
{Foschini} L.,  et~al., 2015, \mn@doi [\aap] {10.1051/0004-6361/201424972},
  \href {https://ui.adsabs.harvard.edu/abs/2015A&A...575A..13F} {575, A13}

\bibitem[\protect\citeauthoryear{{Gallo}}{{Gallo}}{2018}]{Gallo2018}
{Gallo} L.,  2018, in Revisiting Narrow-Line Seyfert 1 Galaxies and their Place
  in the Universe. p.~34 (\mn@eprint {arXiv} {1807.09838})

\bibitem[\protect\citeauthoryear{{Gallo}, {Boller}, {Brandt}, {Fabian}  \&
  {Vaughan}}{{Gallo} et~al.}{2004}]{Gallo+2004}
{Gallo} L.~C.,  {Boller} T.,  {Brandt} W.~N.,  {Fabian} A.~C.,   {Vaughan} S.,
  2004, \mn@doi [\aap] {10.1051/0004-6361:20034411}, \href
  {https://ui.adsabs.harvard.edu/abs/2004A&A...417...29G} {417, 29}

\bibitem[\protect\citeauthoryear{{Gallo}, {Brandt}, {Costantini}  \&
  {Fabian}}{{Gallo} et~al.}{2007}]{Gallo+2007}
{Gallo} L.~C.,  {Brandt} W.~N.,  {Costantini} E.,   {Fabian} A.~C.,  2007,
  \mn@doi [\mnras] {10.1111/j.1365-2966.2007.11701.x}, \href
  {https://ui.adsabs.harvard.edu/abs/2007MNRAS.377.1375G} {377, 1375}

\bibitem[\protect\citeauthoryear{{Gallo}, {Blue}, {Grupe}, {Komossa}  \&
  {Wilkins}}{{Gallo} et~al.}{2018}]{Gallo+2018}
{Gallo} L.~C.,  {Blue} D.~M.,  {Grupe} D.,  {Komossa} S.,   {Wilkins} D.~R.,
  2018, \mn@doi [\mnras] {10.1093/mnras/sty1134}, \href
  {https://ui.adsabs.harvard.edu/abs/2018MNRAS.478.2557G} {478, 2557}

\bibitem[\protect\citeauthoryear{{Garc{\'\i}a} \& {Kallman}}{{Garc{\'\i}a} \&
  {Kallman}}{2010}]{GarciaKallman2010}
{Garc{\'\i}a} J.,  {Kallman} T.~R.,  2010, \mn@doi [\apj]
  {10.1088/0004-637X/718/2/695}, \href
  {https://ui.adsabs.harvard.edu/abs/2010ApJ...718..695G} {718, 695}

\bibitem[\protect\citeauthoryear{{Garc{\'\i}a} et~al.,}{{Garc{\'\i}a}
  et~al.}{2014}]{Garcia+2014}
{Garc{\'\i}a} J.,  et~al., 2014, \mn@doi [\apj] {10.1088/0004-637X/782/2/76},
  \href {https://ui.adsabs.harvard.edu/abs/2014ApJ...782...76G} {782, 76}

\bibitem[\protect\citeauthoryear{{Giroletti}, {Panessa}, {Longinotti},
  {Krongold}, {Guainazzi}, {Costantini}  \& {Santos-Lleo}}{{Giroletti}
  et~al.}{2017}]{Giroletti+2017}
{Giroletti} M.,  {Panessa} F.,  {Longinotti} A.~L.,  {Krongold} Y.,
  {Guainazzi} M.,  {Costantini} E.,   {Santos-Lleo} M.,  2017, \mn@doi [\aap]
  {10.1051/0004-6361/201630161}, \href
  {https://ui.adsabs.harvard.edu/abs/2017A&A...600A..87G} {600, A87}

\bibitem[\protect\citeauthoryear{{Gofford}, {Reeves}, {Tombesi}, {Braito},
  {Turner}, {Miller}  \& {Cappi}}{{Gofford} et~al.}{2013}]{Gofford+2013}
{Gofford} J.,  {Reeves} J.~N.,  {Tombesi} F.,  {Braito} V.,  {Turner} T.~J.,
  {Miller} L.,   {Cappi} M.,  2013, \mn@doi [\mnras] {10.1093/mnras/sts481},
  \href {https://ui.adsabs.harvard.edu/abs/2013MNRAS.430...60G} {430, 60}

\bibitem[\protect\citeauthoryear{{Gonz{\'a}lez-Mart{\'\i}n} \&
  {Vaughan}}{{Gonz{\'a}lez-Mart{\'\i}n} \&
  {Vaughan}}{2012}]{GonzalezMartinVaughan2012}
{Gonz{\'a}lez-Mart{\'\i}n} O.,  {Vaughan} S.,  2012, \mn@doi [\aap]
  {10.1051/0004-6361/201219008}, \href
  {https://ui.adsabs.harvard.edu/abs/2012A&A...544A..80G} {544, A80}

\bibitem[\protect\citeauthoryear{{Gonzalez}, {Wilkins}  \& {Gallo}}{{Gonzalez}
  et~al.}{2017}]{Gonzalez+2017}
{Gonzalez} A.~G.,  {Wilkins} D.~R.,   {Gallo} L.~C.,  2017, \mn@doi [\mnras]
  {10.1093/mnras/stx2080}, \href
  {https://ui.adsabs.harvard.edu/abs/2017MNRAS.472.1932G} {472, 1932}

\bibitem[\protect\citeauthoryear{{Goodrich}}{{Goodrich}}{1989}]{Goodrich1989}
{Goodrich} R.~W.,  1989, \mn@doi [\apj] {10.1086/167586}, \href
  {https://ui.adsabs.harvard.edu/abs/1989ApJ...342..224G} {342, 224}

\bibitem[\protect\citeauthoryear{{Igo} et~al.,}{{Igo} et~al.}{2020}]{Igo+2020}
{Igo} Z.,  et~al., 2020, \mn@doi [\mnras] {10.1093/mnras/staa265}, \href
  {https://ui.adsabs.harvard.edu/abs/2020MNRAS.493.1088I} {493, 1088}

\bibitem[\protect\citeauthoryear{{Jansen} et~al.,}{{Jansen}
  et~al.}{2001}]{Jansen+2001}
{Jansen} F.,  et~al., 2001, \mn@doi [\aap] {10.1051/0004-6361:20000036}, \href
  {http://adsabs.harvard.edu/abs/2001A%26A...365L...1J} {365, L1}

\bibitem[\protect\citeauthoryear{{Jiang} et~al.,}{{Jiang}
  et~al.}{2019}]{Jiang+2019}
{Jiang} J.,  et~al., 2019, \mn@doi [\mnras] {10.1093/mnras/stz2326}, \href
  {https://ui.adsabs.harvard.edu/abs/2019MNRAS.489.3436J} {489, 3436}

\bibitem[\protect\citeauthoryear{{Kaastra}}{{Kaastra}}{2017}]{Kaastra2017}
{Kaastra} J.~S.,  2017, \mn@doi [\aap] {10.1051/0004-6361/201629319}, \href
  {https://ui.adsabs.harvard.edu/abs/2017A&A...605A..51K} {605, A51}

\bibitem[\protect\citeauthoryear{{Kaastra} \& {Bleeker}}{{Kaastra} \&
  {Bleeker}}{2016}]{KaastraBleeker2016}
{Kaastra} J.~S.,  {Bleeker} J.~A.~M.,  2016, \mn@doi [\aap]
  {10.1051/0004-6361/201527395}, \href
  {https://ui.adsabs.harvard.edu/abs/2016A&A...587A.151K} {587, A151}

\bibitem[\protect\citeauthoryear{{Kaastra}, {Mewe}  \&
  {Nieuwenhuijzen}}{{Kaastra} et~al.}{1996}]{KaastraMeweNieuwenhuijzen1996}
{Kaastra} J.~S.,  {Mewe} R.,   {Nieuwenhuijzen} H.,  1996, in UV and X-ray
  Spectroscopy of Astrophysical and Laboratory Plasmas. pp 411--414

\bibitem[\protect\citeauthoryear{{Kallman}}{{Kallman}}{1995}]{Kallman1995}
{Kallman} T.~R.,  1995, \mn@doi [\apj] {10.1086/176608}, \href
  {https://ui.adsabs.harvard.edu/abs/1995ApJ...455..603K} {455, 603}

\bibitem[\protect\citeauthoryear{{Kara}, {Fabian}, {Cackett}, {Steiner},
  {Uttley}, {Wilkins}  \& {Zoghbi}}{{Kara} et~al.}{2013}]{Kara+2013}
{Kara} E.,  {Fabian} A.~C.,  {Cackett} E.~M.,  {Steiner} J.~F.,  {Uttley} P.,
  {Wilkins} D.~R.,   {Zoghbi} A.,  2013, \mn@doi [\mnras]
  {10.1093/mnras/sts155}, \href
  {https://ui.adsabs.harvard.edu/abs/2013MNRAS.428.2795K} {428, 2795}

\bibitem[\protect\citeauthoryear{{Kara}, {Alston}, {Fabian}, {Cackett},
  {Uttley}, {Reynolds}  \& {Zoghbi}}{{Kara} et~al.}{2016}]{Kara+2016}
{Kara} E.,  {Alston} W.~N.,  {Fabian} A.~C.,  {Cackett} E.~M.,  {Uttley} P.,
  {Reynolds} C.~S.,   {Zoghbi} A.,  2016, \mn@doi [\mnras]
  {10.1093/mnras/stw1695}, \href
  {https://ui.adsabs.harvard.edu/abs/2016MNRAS.462..511K} {462, 511}

\bibitem[\protect\citeauthoryear{{Komossa} \& {Bade}}{{Komossa} \&
  {Bade}}{1998}]{KomossaBade1998}
{Komossa} S.,  {Bade} N.,  1998, \aap, \href
  {https://ui.adsabs.harvard.edu/abs/1998A&A...331L..49K} {331, L49}

\bibitem[\protect\citeauthoryear{{Komossa}, {Voges}, {Xu}, {Mathur}, {Adorf},
  {Lemson}, {Duschl}  \& {Grupe}}{{Komossa} et~al.}{2006}]{Komossa+2006}
{Komossa} S.,  {Voges} W.,  {Xu} D.,  {Mathur} S.,  {Adorf} H.-M.,  {Lemson}
  G.,  {Duschl} W.~J.,   {Grupe} D.,  2006, \mn@doi [\aj] {10.1086/505043},
  \href {https://ui.adsabs.harvard.edu/abs/2006AJ....132..531K} {132, 531}

\bibitem[\protect\citeauthoryear{{Komossa} et~al.,}{{Komossa}
  et~al.}{2017}]{Komossa+2017}
{Komossa} S.,  et~al., 2017, in {Gomboc} A.,  ed.,  IAU Symposium Vol. 324, New
  Frontiers in Black Hole Astrophysics. pp 168--171,
  \mn@doi{10.1017/S1743921317001648}

\bibitem[\protect\citeauthoryear{{K{\"o}rding}, {Falcke}  \&
  {Corbel}}{{K{\"o}rding} et~al.}{2006}]{Kording+2006}
{K{\"o}rding} E.,  {Falcke} H.,   {Corbel} S.,  2006, \mn@doi [\aap]
  {10.1051/0004-6361:20054144}, \href
  {https://ui.adsabs.harvard.edu/abs/2006A&A...456..439K} {456, 439}

\bibitem[\protect\citeauthoryear{{Kotov}, {Churazov}  \& {Gilfanov}}{{Kotov}
  et~al.}{2001}]{KotovChurazovGilfanov2001}
{Kotov} O.,  {Churazov} E.,   {Gilfanov} M.,  2001, \mn@doi [\mnras]
  {10.1046/j.1365-8711.2001.04769.x}, \href
  {https://ui.adsabs.harvard.edu/abs/2001MNRAS.327..799K} {327, 799}

\bibitem[\protect\citeauthoryear{{L{\"a}hteenm{\"a}ki}
  et~al.,}{{L{\"a}hteenm{\"a}ki} et~al.}{2017}]{Lahteenmaki+2017}
{L{\"a}hteenm{\"a}ki} A.,  et~al., 2017, \mn@doi [\aap]
  {10.1051/0004-6361/201630257}, \href
  {https://ui.adsabs.harvard.edu/abs/2017A&A...603A.100L} {603, A100}

\bibitem[\protect\citeauthoryear{{L{\"a}hteenm{\"a}ki}, {J{\"a}rvel{\"a}},
  {Ramakrishnan}, {Tornikoski}, {Tammi}, {Vera}  \&
  {Chamani}}{{L{\"a}hteenm{\"a}ki} et~al.}{2018}]{Lahteenmaki+2018}
{L{\"a}hteenm{\"a}ki} A.,  {J{\"a}rvel{\"a}} E.,  {Ramakrishnan} V.,
  {Tornikoski} M.,  {Tammi} J.,  {Vera} R.~J.~C.,   {Chamani} W.,  2018,
  \mn@doi [\aap] {10.1051/0004-6361/201833378}, \href
  {https://ui.adsabs.harvard.edu/abs/2018A&A...614L...1L} {614, L1}

\bibitem[\protect\citeauthoryear{{Laor}}{{Laor}}{1991}]{Laor1991}
{Laor} A.,  1991, \mn@doi [\apj] {10.1086/170257}, \href
  {http://adsabs.harvard.edu/abs/1991ApJ...376...90L} {376, 90}

\bibitem[\protect\citeauthoryear{{Leighly}}{{Leighly}}{1999a}]{LeighlyTiming1999}
{Leighly} K.~M.,  1999a, \mn@doi [\apjs] {10.1086/313277}, \href
  {https://ui.adsabs.harvard.edu/abs/1999ApJS..125..297L} {125, 297}

\bibitem[\protect\citeauthoryear{{Leighly}}{{Leighly}}{1999b}]{LeighlySpectral1999}
{Leighly} K.~M.,  1999b, \mn@doi [\apjs] {10.1086/313287}, \href
  {https://ui.adsabs.harvard.edu/abs/1999ApJS..125..317L} {125, 317}

\bibitem[\protect\citeauthoryear{{Leighly}, {Kay}, {Wills}, {Wills}  \&
  {Grupe}}{{Leighly} et~al.}{1997}]{Leighly+1997}
{Leighly} K.~M.,  {Kay} L.~E.,  {Wills} B.~J.,  {Wills} D.,   {Grupe} D.,
  1997, \mn@doi [\apjl] {10.1086/316793}, \href
  {https://ui.adsabs.harvard.edu/abs/1997ApJ...489L.137L} {489, L137}

\bibitem[\protect\citeauthoryear{{Lohfink} et~al.,}{{Lohfink}
  et~al.}{2013}]{Lohfink+2013}
{Lohfink} A.~M.,  et~al., 2013, \mn@doi [\apj] {10.1088/0004-637X/772/2/83},
  \href {https://ui.adsabs.harvard.edu/abs/2013ApJ...772...83L} {772, 83}

\bibitem[\protect\citeauthoryear{{Longinotti}, {Krongold}, {Guainazzi},
  {Giroletti}, {Panessa}, {Costantini}, {Santos-Lleo}  \&
  {Rodriguez-Pascual}}{{Longinotti} et~al.}{2015}]{Longinotti+2015}
{Longinotti} A.~L.,  {Krongold} Y.,  {Guainazzi} M.,  {Giroletti} M.,
  {Panessa} F.,  {Costantini} E.,  {Santos-Lleo} M.,   {Rodriguez-Pascual} P.,
  2015, \mn@doi [\apjl] {10.1088/2041-8205/813/2/L39}, \href
  {https://ui.adsabs.harvard.edu/abs/2015ApJ...813L..39L} {813, L39}

\bibitem[\protect\citeauthoryear{{Lyubarskii}}{{Lyubarskii}}{1997}]{Lyubarskii1997}
{Lyubarskii} Y.~E.,  1997, \mn@doi [\mnras] {10.1093/mnras/292.3.679}, \href
  {https://ui.adsabs.harvard.edu/abs/1997MNRAS.292..679L} {292, 679}

\bibitem[\protect\citeauthoryear{{Mallick}, {Dewangan}, {Gandhi}, {Misra}  \&
  {Kembhavi}}{{Mallick} et~al.}{2016}]{Mallick+2016}
{Mallick} L.,  {Dewangan} G.~C.,  {Gandhi} P.,  {Misra} R.,   {Kembhavi} A.~K.,
   2016, \mn@doi [\mnras] {10.1093/mnras/stw1073}, \href
  {https://ui.adsabs.harvard.edu/abs/2016MNRAS.460.1705M} {460, 1705}

\bibitem[\protect\citeauthoryear{{Markowitz}, {Edelson}  \&
  {Vaughan}}{{Markowitz} et~al.}{2003}]{MarkowitzEdelsonVaughan2003}
{Markowitz} A.,  {Edelson} R.,   {Vaughan} S.,  2003, \mn@doi [\apj]
  {10.1086/379103}, \href
  {https://ui.adsabs.harvard.edu/abs/2003ApJ...598..935M} {598, 935}

\bibitem[\protect\citeauthoryear{{McHardy}, {Koerding}, {Knigge}, {Uttley}  \&
  {Fender}}{{McHardy} et~al.}{2006}]{McHardy+2006}
{McHardy} I.~M.,  {Koerding} E.,  {Knigge} C.,  {Uttley} P.,   {Fender} R.~P.,
  2006, \mn@doi [\nat] {10.1038/nature05389}, \href
  {https://ui.adsabs.harvard.edu/abs/2006Natur.444..730M} {444, 730}

\bibitem[\protect\citeauthoryear{{Merloni}, {Heinz}  \& {di Matteo}}{{Merloni}
  et~al.}{2003}]{Merloni+2003}
{Merloni} A.,  {Heinz} S.,   {di Matteo} T.,  2003, \mn@doi [\mnras]
  {10.1046/j.1365-2966.2003.07017.x}, \href
  {https://ui.adsabs.harvard.edu/abs/2003MNRAS.345.1057M} {345, 1057}

\bibitem[\protect\citeauthoryear{{Niko{\l}ajuk}, {Czerny}  \&
  {Gurynowicz}}{{Niko{\l}ajuk} et~al.}{2009}]{Nikolajuk+2009}
{Niko{\l}ajuk} M.,  {Czerny} B.,   {Gurynowicz} P.,  2009, \mn@doi [\mnras]
  {10.1111/j.1365-2966.2009.14478.x}, \href
  {https://ui.adsabs.harvard.edu/abs/2009MNRAS.394.2141N} {394, 2141}

\bibitem[\protect\citeauthoryear{{Osterbrock} \& {Pogge}}{{Osterbrock} \&
  {Pogge}}{1985}]{OsterbrockPogge1985}
{Osterbrock} D.~E.,  {Pogge} R.~W.,  1985, \mn@doi [\apj] {10.1086/163513},
  \href {https://ui.adsabs.harvard.edu/abs/1985ApJ...297..166O} {297, 166}

\bibitem[\protect\citeauthoryear{{Peterson}, {McHardy}  \& {Wilkes}}{{Peterson}
  et~al.}{2000}]{Peterson+2000}
{Peterson} B.~M.,  {McHardy} I.~M.,   {Wilkes} B.~J.,  2000, \mn@doi [\nar]
  {10.1016/S1387-6473(00)00086-5}, \href
  {https://ui.adsabs.harvard.edu/abs/2000NewAR..44..491P} {44, 491}

\bibitem[\protect\citeauthoryear{{Petrucci}, {Merloni}, {Fabian}, {Haardt}  \&
  {Gallo}}{{Petrucci} et~al.}{2001}]{Petrucci+2001}
{Petrucci} P.~O.,  {Merloni} A.,  {Fabian} A.,  {Haardt} F.,   {Gallo} E.,
  2001, \mn@doi [\mnras] {10.1046/j.1365-8711.2001.04897.x}, \href
  {https://ui.adsabs.harvard.edu/abs/2001MNRAS.328..501P} {328, 501}

\bibitem[\protect\citeauthoryear{{Petrucci}, {Ursini}, {De Rosa}, {Bianchi},
  {Cappi}, {Matt}, {Dadina}  \& {Malzac}}{{Petrucci}
  et~al.}{2018}]{Petrucci+2018}
{Petrucci} P.~O.,  {Ursini} F.,  {De Rosa} A.,  {Bianchi} S.,  {Cappi} M.,
  {Matt} G.,  {Dadina} M.,   {Malzac} J.,  2018, \mn@doi [\aap]
  {10.1051/0004-6361/201731580}, \href
  {https://ui.adsabs.harvard.edu/abs/2018A&A...611A..59P} {611, A59}

\bibitem[\protect\citeauthoryear{{Sanfrutos}, {Longinotti}, {Krongold},
  {Guainazzi}  \& {Panessa}}{{Sanfrutos} et~al.}{2018}]{Sanfrutos+2018}
{Sanfrutos} M.,  {Longinotti} A.~L.,  {Krongold} Y.,  {Guainazzi} M.,
  {Panessa} F.,  2018, \mn@doi [The Astrophysical Journal]
  {10.3847/1538-4357/aae923}, \href
  {https://ui.adsabs.harvard.edu/abs/2018ApJ...868..111S} {868, 111}

\bibitem[\protect\citeauthoryear{{Shakura} \& {Sunyaev}}{{Shakura} \&
  {Sunyaev}}{1973}]{ShakuraSunyaev1973}
{Shakura} N.~I.,  {Sunyaev} R.~A.,  1973, \aap, \href
  {https://ui.adsabs.harvard.edu/abs/1973A&A....24..337S} {500, 33}

\bibitem[\protect\citeauthoryear{{Singh}, {Meintjes}, {Ramamonjisoa}  \&
  {Tolamatti}}{{Singh} et~al.}{2019}]{Singh+2019}
{Singh} K.,  {Meintjes} P.,  {Ramamonjisoa} F.,   {Tolamatti} A.,  2019,
  \mn@doi [\na] {10.1016/j.newast.2019.101278}, \href
  {https://ui.adsabs.harvard.edu/abs/2019NewA...7301278S} {73, 101278}

\bibitem[\protect\citeauthoryear{{Snellen}, {Mack}, {Schilizzi}  \&
  {Tschager}}{{Snellen} et~al.}{2004}]{Snellen+2004}
{Snellen} I.~A.~G.,  {Mack} K.~H.,  {Schilizzi} R.~T.,   {Tschager} W.,  2004,
  \mn@doi [\mnras] {10.1111/j.1365-2966.2004.07337.x}, \href
  {https://ui.adsabs.harvard.edu/abs/2004MNRAS.348..227S} {348, 227}

\bibitem[\protect\citeauthoryear{{Steenbrugge}, {Kaastra}, {de Vries}  \&
  {Edelson}}{{Steenbrugge} et~al.}{2003}]{Steenbrugge+2003}
{Steenbrugge} K.~C.,  {Kaastra} J.~S.,  {de Vries} C.~P.,   {Edelson} R.,
  2003, \mn@doi [\aap] {10.1051/0004-6361:20030261}, \href
  {https://ui.adsabs.harvard.edu/abs/2003A&A...402..477S} {402, 477}

\bibitem[\protect\citeauthoryear{{Str{\"u}der} et~al.,}{{Str{\"u}der}
  et~al.}{2001}]{Struder+2001}
{Str{\"u}der} L.,  et~al., 2001, \mn@doi [\aap] {10.1051/0004-6361:20000066},
  \href {http://adsabs.harvard.edu/abs/2001A%26A...365L..18S} {365, L18}

\bibitem[\protect\citeauthoryear{{Taylor}, {Uttley}  \& {McHardy}}{{Taylor}
  et~al.}{2003}]{Taylor+2003}
{Taylor} R.~D.,  {Uttley} P.,   {McHardy} I.~M.,  2003, \mn@doi [\mnras]
  {10.1046/j.1365-8711.2003.06742.x}, \href
  {https://ui.adsabs.harvard.edu/abs/2003MNRAS.342L..31T} {342, L31}

\bibitem[\protect\citeauthoryear{{Titarchuk}}{{Titarchuk}}{1994}]{Titarchuk+1994}
{Titarchuk} L.,  1994, \mn@doi [\apj] {10.1086/174760}, \href
  {https://ui.adsabs.harvard.edu/abs/1994ApJ...434..570T} {434, 570}

\bibitem[\protect\citeauthoryear{{Tombesi}, {Cappi}, {Reeves}, {Palumbo},
  {Yaqoob}, {Braito}  \& {Dadina}}{{Tombesi} et~al.}{2010}]{Tombesi+2010}
{Tombesi} F.,  {Cappi} M.,  {Reeves} J.~N.,  {Palumbo} G.~G.~C.,  {Yaqoob} T.,
  {Braito} V.,   {Dadina} M.,  2010, \mn@doi [\aap]
  {10.1051/0004-6361/200913440}, \href
  {https://ui.adsabs.harvard.edu/abs/2010A&A...521A..57T} {521, A57}

\bibitem[\protect\citeauthoryear{{Uttley}, {McHardy}  \& {Vaughan}}{{Uttley}
  et~al.}{2005}]{UttleyMcHardyVaughan2005}
{Uttley} P.,  {McHardy} I.~M.,   {Vaughan} S.,  2005, \mn@doi [\mnras]
  {10.1111/j.1365-2966.2005.08886.x}, \href
  {https://ui.adsabs.harvard.edu/abs/2005MNRAS.359..345U} {359, 345}

\bibitem[\protect\citeauthoryear{{Uttley}, {Cackett}, {Fabian}, {Kara}  \&
  {Wilkins}}{{Uttley} et~al.}{2014}]{Uttley+2014}
{Uttley} P.,  {Cackett} E.~M.,  {Fabian} A.~C.,  {Kara} E.,   {Wilkins} D.~R.,
  2014, \mn@doi [Astronomy and Astrophysics Review]
  {10.1007/s00159-014-0072-0}, \href
  {https://ui.adsabs.harvard.edu/\#abs/2014A&ARv..22...72U} {22, 72}

\bibitem[\protect\citeauthoryear{{Vaughan}}{{Vaughan}}{2010}]{Vaughan2010}
{Vaughan} S.,  2010, \mn@doi [\mnras] {10.1111/j.1365-2966.2009.15868.x}, \href
  {https://ui.adsabs.harvard.edu/abs/2010MNRAS.402..307V} {402, 307}

\bibitem[\protect\citeauthoryear{{Vaughan}, {Edelson}, {Warwick}  \&
  {Uttley}}{{Vaughan} et~al.}{2003}]{Vaughan+2003}
{Vaughan} S.,  {Edelson} R.,  {Warwick} R.~S.,   {Uttley} P.,  2003, \mn@doi
  [\mnras] {10.1046/j.1365-2966.2003.07042.x}, \href
  {https://ui.adsabs.harvard.edu/abs/2003MNRAS.345.1271V} {345, 1271}

\bibitem[\protect\citeauthoryear{{Wilkins} \& {Fabian}}{{Wilkins} \&
  {Fabian}}{2012}]{WilkinsFabian2012}
{Wilkins} D.~R.,  {Fabian} A.~C.,  2012, \mn@doi [\mnras]
  {10.1111/j.1365-2966.2012.21308.x}, \href
  {https://ui.adsabs.harvard.edu/abs/2012MNRAS.424.1284W} {424, 1284}

\bibitem[\protect\citeauthoryear{{Wilkins} \& {Gallo}}{{Wilkins} \&
  {Gallo}}{2015a}]{WilkinsGalloCOMP2015}
{Wilkins} D.~R.,  {Gallo} L.~C.,  2015a, \mn@doi [\mnras]
  {10.1093/mnras/stu2524}, \href
  {https://ui.adsabs.harvard.edu/abs/2015MNRAS.448..703W} {448, 703}

\bibitem[\protect\citeauthoryear{{Wilkins} \& {Gallo}}{{Wilkins} \&
  {Gallo}}{2015b}]{WilkinsGallo2015}
{Wilkins} D.~R.,  {Gallo} L.~C.,  2015b, \mn@doi [\mnras]
  {10.1093/mnras/stv162}, \href
  {https://ui.adsabs.harvard.edu/abs/2015MNRAS.449..129W} {449, 129}

\bibitem[\protect\citeauthoryear{{Wilkins}, {Gallo}, {Silva}, {Costantini},
  {Brandt}  \& {Kriss}}{{Wilkins} et~al.}{2017}]{Wilkins+2017}
{Wilkins} D.~R.,  {Gallo} L.~C.,  {Silva} C.~V.,  {Costantini} E.,  {Brandt}
  W.~N.,   {Kriss} G.~A.,  2017, \mn@doi [\mnras] {10.1093/mnras/stx1814},
  \href {https://ui.adsabs.harvard.edu/abs/2017MNRAS.471.4436W} {471, 4436}

\bibitem[\protect\citeauthoryear{{Willingale}, {Starling}, {Beardmore},
  {Tanvir}  \& {O'Brien}}{{Willingale} et~al.}{2013}]{Willingale+2013}
{Willingale} R.,  {Starling} R.~L.~C.,  {Beardmore} A.~P.,  {Tanvir} N.~R.,
  {O'Brien} P.~T.,  2013, \mn@doi [\mnras] {10.1093/mnras/stt175}, \href
  {https://ui.adsabs.harvard.edu/abs/2013MNRAS.431..394W} {431, 394}

\bibitem[\protect\citeauthoryear{{Wilms}, {Allen}  \& {McCray}}{{Wilms}
  et~al.}{2000}]{Wilms+2000}
{Wilms} J.,  {Allen} A.,   {McCray} R.,  2000, \mn@doi [\apj] {10.1086/317016},
  \href {https://ui.adsabs.harvard.edu/abs/2000ApJ...542..914W} {542, 914}

\bibitem[\protect\citeauthoryear{{Zhang}, {Gupta}, {Gaur}, {Wiita}, {An}, {Gu},
  {Hu}  \& {Xu}}{{Zhang} et~al.}{2019}]{Zhang+2019}
{Zhang} Z.,  {Gupta} A.~C.,  {Gaur} H.,  {Wiita} P.~J.,  {An} T.,  {Gu} M.,
  {Hu} D.,   {Xu} H.,  2019, \mn@doi [\apj] {10.3847/1538-4357/ab3f3a}, \href
  {https://ui.adsabs.harvard.edu/abs/2019ApJ...884..125Z} {884, 125}

\bibitem[\protect\citeauthoryear{{Zoghbi}, {Fabian}, {Uttley}, {Miniutti},
  {Gallo}, {Reynolds}, {Miller}  \& {Ponti}}{{Zoghbi}
  et~al.}{2010}]{Zoghbi+2010}
{Zoghbi} A.,  {Fabian} A.~C.,  {Uttley} P.,  {Miniutti} G.,  {Gallo} L.~C.,
  {Reynolds} C.~S.,  {Miller} J.~M.,   {Ponti} G.,  2010, \mn@doi [\mnras]
  {10.1111/j.1365-2966.2009.15816.x}, \href
  {https://ui.adsabs.harvard.edu/abs/2010MNRAS.401.2419Z} {401, 2419}

\makeatother
\end{thebibliography}

\bsp	
\label{lastpage}
\end{document}